\documentclass[10pt,journal,final,letterpaper,twocolumn,twoside]{IEEEtran} %For twocolumn uncomment

\usepackage{amsmath,amsfonts,bm,amssymb,psfrag,ifthen,color,subfigure}
\usepackage{mathrsfs}
\usepackage[font=footnotesize]{caption}

\usepackage[dvips]{epsfig}
\usepackage[dvips]{graphicx}
\usepackage{amsmath,amsfonts,bm,amssymb,psfrag,ifthen,color,subfigure}
\allowdisplaybreaks[0]
\usepackage{algpseudocode}
\usepackage{algorithm}
\usepackage{algorithmicx}
\usepackage{ifthen}
\usepackage{setspace}
\usepackage{cite}
\usepackage{url}
\usepackage{longtable}
\usepackage{stfloats}  % Written by Sigitas Tolusis
\usepackage{graphics,booktabs,color,subfigure}
\usepackage{float}

\usepackage{listings}

\floatname{algorithm}{Algorithm}

\newcounter{mytempeqncnt}

\begin{document}
\title{Spectral  and Energy Efficiency of ACO-OFDM in Visible
Light Communication Systems}
 \author{Shuai Ma, Ruixin Yang, Xiong Deng,~\IEEEmembership{Member,~IEEE}, Xintong Ling,~\IEEEmembership{Member,~IEEE}, Xun Zhang, Fuhui Zhou, Shiyin Li, and Derrick Wing Kwan Ng,~\IEEEmembership{Fellow,~IEEE}
 \thanks{S. Ma, R. Yang and S. Li   are with the School of Information and Control   Engineering, China
University of Mining and Technology, Xuzhou, 221116,
China. (e-mail:mashuai001@cumt.edu.cn; ray.young@cumt.edu.cn; lishiyin@cumt.edu.cn).}
\thanks{X. Deng is with the Department of Electrical Engineering, Eindhoven University of Technology
(TU/e), Eindhoven, NL. (e-mail:  X.Deng@tue.nl).}
 \thanks{X. Ling is with the National Mobile Communications Research Laboratory, Southeast University,
 and the Purple Mountain Laboratories, Nanjing, China. (e-mail: xtling@seu.edu.cn). }
  \thanks{X. Zhang is with Institut Suprieur dElectronique de Paris, ISEP Paris, France. (e-mail: xun.zhang@isep.fr).}
 \thanks{F. Zhou is with the College of Electronic and Information Engineering,
 Nanjing University of Aeronautics and Astronautics, Nanjing, 210000,  China. He is also with Key Laboratory of Dynamic Cognitive System of Electromagnetic Spectrum Space, Nanjing University of Aeronautics and Astronautics. (e-mail: zhoufuhui@ieee.org).}
 \thanks{D. W. K. Ng is with the School of Electrical Engineering and Telecommunications,
University of New SouthWales, Sydney, NSW 2052, Australia (e-mail:w.k.ng@unsw.edu.au).}

}

\maketitle
\begin{abstract}

In this paper, we study  the spectral efficiency (SE) and energy efficiency (EE)   of asymmetrically clipped optical orthogonal frequency division multiplexing (ACO-OFDM)  for visible light communication (VLC). Firstly, we derive the achievable rates for  Gaussian distributions inputs and practical finite-alphabet inputs. Then, we investigate  the  SE  maximization problems  subject to both  the total transmit   power constraint and the average optical power constraint with the above two inputs, respectively. By exploiting   the relationship between the mutual information and the minimum mean-squared error,  an optimal power allocation scheme is proposed  to maximize the SE with  finite-alphabet inputs. To reduce the computational complexity of the  power allocation scheme,   we derive a  closed-form  lower bound of the SE.  Also,   considering  the quality of service, we further tackle the non-convex EE  maximization problems of ACO-OFDM   with the two inputs,  respectively. The problems are solved by the proposed Dinkelbach-type iterative algorithm. In each iteration, the interior point algorithm is applied to obtain the optimal power allocation.The performance of the proposed  power allocation schemes for the SE and EE maximization are  validated through numerical analysis.

\end{abstract}


\IEEEpeerreviewmaketitle

\section{Introduction}

Traditional radio frequency (RF) communications are facing the problem of spectrum crunch  because of the
exponential increase in the demand for wireless  data traffic \cite{Hanzo,Wong2017}. Besides,
the tremendous wireless  devices consume more than $3\%$ of the global
energy\cite{Fettweis,Li_Xu}, and lead to about $5\%$ of the total ${\text{C}}{{\text{O}}_{\text{2}}}$
 emissions
worldwide by 2020 \cite{Buzzi,Ismail_Zhuang,Wu_Schober}.
Therefore, both  spectral  and
 energy resources are severely limited for next generation wireless communications.
Facilitated by the
 low-cost and widely installed lighting infrastructure with light emitting diodes (LEDs),
 visible light communication (VLC)
  has emerged as a promising green indoor communication solution enabling simultaneous illumination
and wireless data transmission.
Owing  to its   inherent advantages, such as abundant license-free
spectrum, high security, and  no interference to existing
RF-based systems, VLC systems are
    a compelling supplementary to  RF
systems   for realizing high-speed   wireless data transmissions.

Despite the promising gains brought VLC technologies, serious  inter-symbol interference (ISI)
creates a system performance bottleneck due to the existence of   multipath
  in high-data rate VLC  systems.
 In practice, orthogonal frequency-division multiplexing (OFDM)\cite{Armstrong2} is  an  effective solution to trackle   ISI in RF-based systems.
However, as VLC systems exploit  intensity
modulation and direct detection (IM/DD) schemes for communication,   information of VLC  is represented   by light intensity  and thus   transmitted signals should be real-valued and nonnegative. Thus, conventional RF-based OFDM  techniques  cannot directly apply to VLC systems.
To mitigate the ISI issue,
asymmetrically clipped optical OFDM (ACO-OFDM) \cite{Armstrong3,Li2},  direct current biased optical OFDM (DCO-OFDM) \cite{Mardanikorani_22,Ling2,Chen2016OSA},  and Unipolar OFDM (U-OFDM)\cite{Dobroslav2012VTC}  have been  proposed  for VLC systems.
     To generate nonnegative   transmitted signals, ACO-OFDM eliminates
       the negative component  of   signals, while DCO-OFDM adds
      a direct current (DC) bias and then clips  the  negative parts of  signals by setting them to zero\cite{Mardanikorani}.
    Moreover,  ACO-OFDM transmits data  symbols only via
     odd indexed subcarriers, whereas   DCO-OFDM transmits data  symbols exploiting  all the subcarriers.
      Compared with DCO-OFDM, ACO-OFDM can generally achieve a lower bit-error-rate (BER) for identical QAM modulation orders such as in\cite{Dimitrov2012,Dissanayake}.
     However, due to  only half of the subcarriers to carry information, the  spectral efficiency (SE) of ACO-OFDM
 is generally lower than that of DCO-OFDM from moderate to high signal-to-noise ratio (SNR) \cite{Mazahir_ICC}, and the SE of U-OFDM is similar to that of ACO-OFDM\cite{ZhaochengWCMC2017}.

 Recently, various power allocation schemes have been proposed  to improve the SE of  ACO-OFDM VLC systems. For example, under average optical power constraint, the conventional
        water-filling power allocation scheme  can improve the information rate of  ACO-OFDM substantially  \cite{Li2}.
  In \cite{Mazahir}, the achievable rates of ACO-OFDM and filtered ACO-OFDM (FACO-OFDM) were analyzed with both  optical
power and bandwidth constraints. Besides,
by taking into account  both  average optical
power and dynamic optical power constraints,
   both the error vector magnitude (EVM) and
achievable data rates of the DCO-OFDM
and ACO-OFDM systems were analyzed in \cite{Zhenhua}    showing that ACO-OFDM can achieve
 the lower bound of  the EVM.
 In \cite{Mardling_ICC}, two upper bounds of channel capacity for the intensity modulated direct detection (IM/DD) optical communication systems were derived based on an exponential input distribution and clipped Gaussian input distribution respectively, and a closed-formed channel capacity of ACO-OFDM. However, all of them only considered the average optical power constraint. Then,  a more detailed description of the problem was given in \cite{Li2} with an electrical power limit or  both optical power limit and input power constraint. For the latter scenario, \cite{Li2} proved  that if the real and imaginary components of each odd frequency IFFT input are independent random variables with circular symmetry, they must follow a zero mean Gaussian distribution and the outputs of the IFFT are strict sense stationary. Then, a closed-form information rate was derived to satisfy above conditions. However, the more general question of the information rate of an ACO-OFDM system only with limited average optical power
     remains an intractable problem.
Also, subject to a given a target BER requirement, adaptive modulation schemes were investigated  in \cite{Wu}
 to  maximize
the SE of DCO-OFDM, ACO-OFDM, and single carrier
frequency-domain equalization (SC-FDE) systems, respectively.
In spite of the fruitful  research in the literature,  the aforementioned
studies were based on the assumption that the input signal follows Gaussian distribution.
Although Gaussian distribution inputs can  achieve the  channel capacity under average electrical power constraints,
 the optimal distribution  with optical power constraints is still unknown.
In fact, practical input signals are often  based on   discrete
constellation schemes, such as pulse amplitude
modulation (PAM), quadrature amplitude modulation (QAM),
and phase shift keying (PSK).
  Applying  power allocation schemes
based on Gaussian distribution inputs to signals with practical finite-alphabet inputs may cause  serious performance loss \cite{Globally}. So far, the
 SE    of ACO-OFDM   with finite-alphabet
inputs     has been
rarely considered in the literature.
Therefore, it is necessary to design an optimal power allocation scheme to
unlock the potential of ACO-OFDM systems.

In addition to improving the SE, achieving   high energy efficiency (EE) is also critical for ACO-OFDM VLC systems,
which is usually defined as a ratio of the achievable  rate to the total power consumption \cite{Ma}.
In fact, the improved SE does not come for free.
In particular, the improvement is always achieved at the expense of increased energy cost. Unfortunately,
most of the aforementioned
 VLC research \cite{Azim} aimed at improving SE, but omitted
 the EE of ACO-OFDM systems.
 Recently, there are some works started focusing on the EE issue in VLC systems.
 By the joint design of the cell structure   and the system level power allocation,
 an amorphous structure of ACO-OFDM VLC systems can  achieve
a   higher EE than that of the  conventional cell structures\cite{Rong}.
To ensure the quality of service (QoS) with affordable energy,
the EE   of  the conventional
and hybrid OFDM-based VLC modulation schemes was investigated in \cite{sun}.
However,    existing studies of ACO-OFDM's EE \cite{Rong,sun}  are based on   Gaussian distribution inputs.
As previously mentioned,    Gaussian distribution inputs  are difficult to
 generate in   practice. Indeed,  practical inputs   are  always  finite-alphabet inputs,  which
     has been less commonly studied in literature.
 Thus,  there is an emerge need for the study of EE of ACO-OFDM VLC systems with finite-alphabet inputs.

In this study, we   propose  the optimal power allocation scheme to maximize the SE and the EE of ACO-OFDM VLC systems  with Gaussian distribution inputs and
 finite-alphabet
inputs, respectively.
    The main contributions of this paper are summarized as follows:
\begin{itemize}
\item We  systematically analyze the signal processing module of a
typical ACO-OFDM VLC system. Based on the frequency domain analysis, we first derive achievable rates of the considered system
admitting finite-alphabet inputs  from the perspective of  practical modulation.
Additionally,
for both cases of the Gaussian distribution inputs and finite-alphabet inputs,
we develop the corresponding
optical power constraints
for   ACO-OFDM VLC systems.

\item Under   both the total transmit power constraint and the optical power constraint, two    optimal power allocation schemes are proposed to maximize the
SE of the ACO-OFDM system with Gaussian distribution inputs and finite-alphabet inputs, respectively.
Specifically, for Gaussian distribution inputs,  we show that the water-filling-based  power allocation scheme can  maximize the SE.
Similarly, for finite-alphabet inputs,  we derive  an optimal power
allocation scheme to achieve the maximum  SE by exploiting the Lagrangian method, Karush-Kuhn-Tucker (KKT) conditions, and
the relationship between the mutual information and the minimum mean-squared error (MMSE)\cite{Lozano}.

\item The   optimal power
allocation scheme  for finite-alphabet inputs lacks  closed-form expressions and involves complicated computations of MMSE.
To reduce the computational complexity, we first derive a closed-form lower bound   for the achievable rate. Then, based on the proposed lower bound, we develop a suboptimal power
allocation scheme to  maximize the SE under   both the total transmit power constraint and the average optical power constraint.

\item We propose an explicit EE expression
with Gaussian distribution  inputs  and finite-alphabet inputs, respectively.
Moreover, under the constraint of  maximum transmission  power and the minimum data rate requirement, the non-convex problem of maximizing EE
is investigated.
This problem is solved by applying the Dinkelbach-type algorithm  and the interior point algorithm.
Finally, the relationship between the SE and the EE of the ACO-OFDM system is unveiled.

\end{itemize}

The rest of this paper is organized as follows.
The system model of ACO-OFDM  is presented in Section II.
The   SE of ACO-OFDM system is shown in Section III.
The EE of ACO-OFDM system  is studied in Section IV  and the simulation results are presented in Section V.
Finally, the conclusions are drawn in Section VI.

\emph{Notations}: Boldfaced lowercase and uppercase letters represent vectors and matrices, respectively.
 Expected value of  a random variable $z$ is denoted by ${\mathbb{E}}\left\{ z \right\}$. ${\left(  \cdot  \right)^ * }$ represents conjugate transformation.
 ${\left[  x  \right]^ + }$ denotes $\max \left\{ {x,0} \right\}$.
 ${\mathop{\rm Re}\nolimits} \left(  \cdot  \right)$
 denotes the real part of its argument.
$\frac{{\partial f\left(  \cdot  \right)}}{{\partial x}}$ represents the partial derivative operation on $x$ of function ${f\left(  \cdot  \right)}$.
Given a  variable $y$, ${\mathbb{E}}\left\{ {z\left| y \right.} \right\}$ represents the
conditional mean of $z$ for given $y$.
${\rm{min}}\left\{ {x,y} \right\}$ represents the minimum value between $x$ and $y$.
$I\left( {X;Y} \right)$ represents the mutual information of $X$ and $Y$.
A complex-valued circularly symmetric Gaussian distribution with   mean   $\mu$ and  variance   ${\sigma ^2}$
is denoted by $\mathcal{CN}\left( {\mu ,{\sigma ^2}} \right)$.
A real-valued Gaussian distribution with   mean   $\mu$ and  variance   ${\sigma ^2}$
is denoted by $\mathcal{N}\left( {\mu ,{\sigma ^2}} \right)$.

\section{System Model}
 \begin{figure}[htbp]
    \centering
    \includegraphics[width=0.5\textwidth]{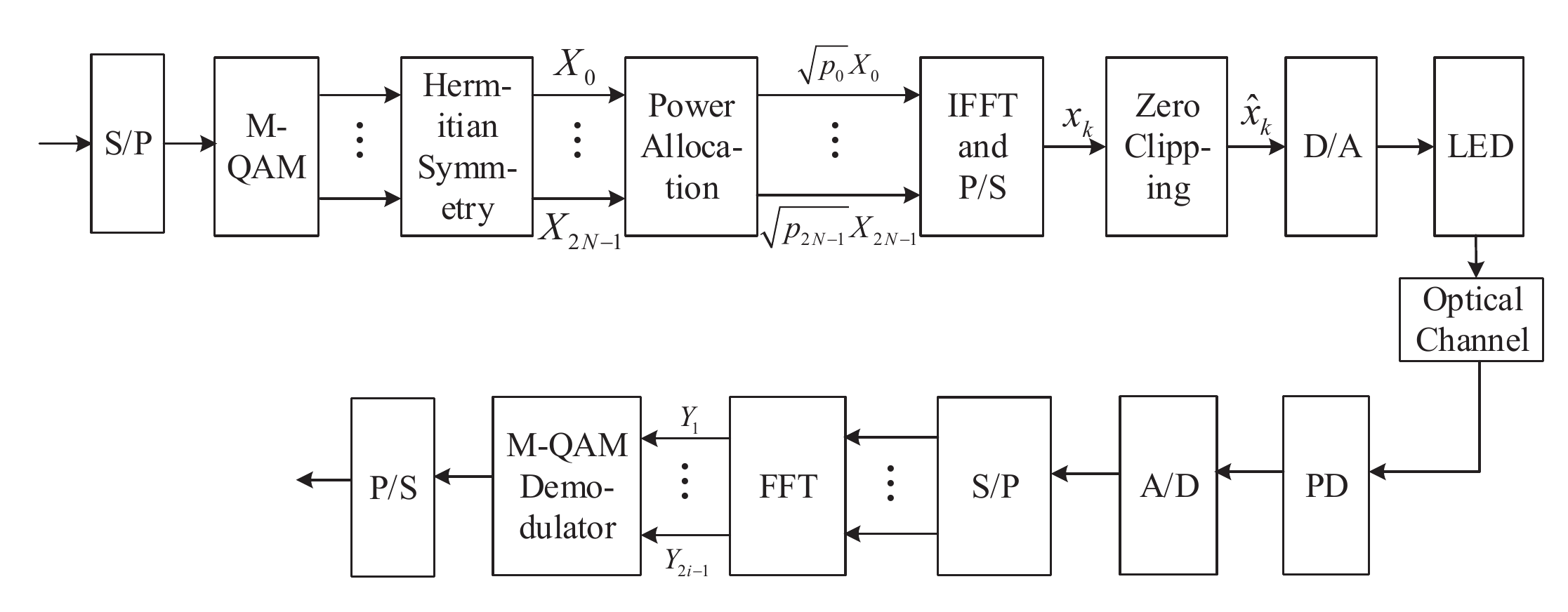}
    \caption{~A block diagram of an ACO-OFDM  VLC system.}
    \label{system}
\end{figure}

Consider an ACO-OFDM VLC system with total $2N$ subcarriers, as shown in Fig.~\ref{system} \cite{Dissanayake},
where the  signals are  only transmitted via the  odd indexed subcarriers.
The information bit stream  is   first converted to parallel sub-streams by a
serial-to-parallel (S/P) converter. Then, they are   modulated by an $M$-QAM scheme.
 After applying the inverse fast Fourier transform (IFFT) and zero clipping on the modulated symbols, signal ${{\hat x}_k}$ is non-negative.
Then, the signal  passes through an  digital-to-analog converter (${\text{D/A}}$), where the digital signal ${{\hat x}_k}$ is converted to an analog signal.
After that, the analog signal  is  emitted through visible light  by an LED.
 In particular, by exploiting  the
IM/DD scheme,  the transmitted information of the VLC system is represented by the signal intensity, which is real and non-negative.
 At the receiver, the received visible
light is transformed into an analog electrical signal by a photo detector (PD) and then converted     to a digital signal by an analog-to-digital converter (${\text{A/D}}$).
   After applying the fast Fourier transform (FFT) on the digitalized signal,
demodulation is performed at a demodulator to convert the received $M$-QAM symbols to bit streams.

\subsection{Signal Model}
In this section, we discuss the mathematical details of the considered system.
At the transmitter, the raw data bit stream is going through the modulation. Let ${X_{k}}$ denotes  the modulated signal on the $k$th subcarrier, and $p_k$ denotes the allocated power on the $k$th subcarrier, $k = 0,...,2N-1$.
Note that to ensure real output values at the IFFT,
the input of the IFFT module  should satisfy  Hermitian symmetry, i.e.,
\begin{align} \label{Herm}
    \begin{cases}
        X_{2i} = X_{2\left(N-i\right)-2} = 0,i = 0, \dots, N/2-1,\\
        X_{2i-1} = X^{*}_{2\left(N-i\right)+1},i = 1, \dots, N/2,
    \end{cases}
\end{align}
where
${X_{2i - 1}}$ is the normalized unit-power input, i.e., $\mathbb{E}\left\{ {{{\left| {{X_{2i - 1}}} \right|}^2}} \right\} = 1$. According to \eqref{Herm},
 the power allocation of subcarriers
should satisfy
\begin{align}   \label{p_i_H}
    \begin{cases}
    p_{2i} = p_{2\left(N-i\right)-2} = 0,i = 0, \dots, N/2-1,\\
    p_{2i-1} = p^{*}_{2\left(N-i\right)+1}\geq 0 ,i = 1, \dots, N/2,
\end{cases}
\end{align}

After the IFFT operation, the time domain signal ${x_k}$ is given  as
 \begin{subequations}
 \begin{align}
 {x_k} = & {\rm{IFFT}}\left\{ \left\{\sqrt {{p _\ell}} {X_\ell}\right\}_{\ell=0}^{2N-1} \right\} \\
   = & \frac{1}{{\sqrt {2N} }}\sum\limits_{\ell = 0}^{2N - 1} {\sqrt {{p_\ell}} {X_\ell}\exp \left( {j\frac{{\pi k\ell}}{N}} \right)}   \\
   =&  \sqrt {\frac{2}{N}}\sum\limits_{i = 1}^{N/2} {\sqrt {{p _{2i - 1}}} {\rm{Re}}\left( {{X_{2i - 1}}\exp \left( {j\frac{{\pi k\left( {2i - 1} \right)}}{N}} \right)} \right)},\nonumber\\
   & \quad  k = 0,...,2N-1.  \label{IFFT_1}
 \end{align}
 \end{subequations}

According to \eqref{IFFT_1},  the obtained   time domain signal satisfies antisymmetry as follows
\begin{align}
{x_\ell} =  - {x_{\ell + N}},\ell = 0,...,N-1.
\end{align}
Since the transmitted signal should be  nonnegative, the negative signals are removed  by the clipping process such that
\begin{align}   \label{zeroclip_ACO}
{{\hat x}_k} = \left\{ {\begin{array}{*{20}{c}}
   {{x_k}}  \\
   0  \\
\end{array}} \right.\begin{array}{*{20}{c}}
   {{x_k} \ge 0};  \\
   {\rm{otherwise} }.  \\
\end{array}
\end{align}

Due to the requirement of the practical system circuit design, the total electrical transmit power should be limited \cite{Li2}. Let $P$   denote the total electrical transmit power, i.e.,  $\sum_{k=0}^{2N-1}\mathbb{E}\left\{ {\hat x_k^2} \right\} \le  {P}$.
Combining the clipping process and
Parseval's theorem \cite{oppenheim1997signals,Mardanikorani_22}, we have  $ \sum_{k = 0}^{2N - 1} {{p_k}}  = \sum_{k = 0}^{2N - 1} \mathbb{E}\left\{ {x_k^2} \right\}=2  \sum_{k = 0}^{2N - 1} \mathbb{E}\left\{ {\hat x_k^2} \right\}$. Based on \eqref{p_i_H},  the electrical transmit power constraint can be   rewritten as
 $\sum_{i = 1}^{N/2} {{p_{2i - 1}}}  \le P$.
For the consideration of human eye safety, the   optical power of VLC signals is generally restricted \cite{BoBai,KYing,Rajagopal,GaoJTOD,WangHu}.
Let  ${P_o}$ represent the maximum  optical power threshold.
The average optical power should satisfy\footnote{Due to different  distributions, the signals with the same  optical power may have  different  electrical powers.}
 \begin{align}  \label{dimming_aco}
\mathbb{E}\left\{ {{\hat x}_k} \right\} \le  {P_o}.
\end{align}
According to the definition of variance,
it is easy to verify that the sum of average electrical power $P$
is larger than  $P_o^2$.

\subsection{Channel Model}
Generally,  the   VLC  channel
is characterized by  a line-of-sight (LOS) link along with   multiple reflections of the light from
 surrounding objects, such as walls,  floor, and windows. In this study, we adopt the commonly used frequency-domain
 VLC channel model \cite{Schulze}, which is not restricted to a finite order of reflections.

Let $H_k$ denotes the    channel gain of     the $k$th subcarrier,  which includes    both the LOS link
and the diffuse links as follows
\begin{align} \label{channel_model}
{H_k} = {H_{{\rm{L}},k}} + {H_{{\rm{D}},k}},
\end{align}
where  ${H_{{\rm{L,}}k}}$
 is the  gain of the LOS link and
 ${H_{{\rm{D,}}i}}$
 is the gain of the diffuse links, {$k = 0,...,2N-1$}.

The LOS link {${H_{{\rm{L,}}k}}$} is expressed as
\begin{align}\label{channel_model_direct}
{H_{{\rm{L,}}k}} = {g _{\rm{L}}}{e^{ - j2\pi f_k \tau }},
\end{align}
where ${g _{\rm{L}}}$ is the  generalized Lambertian radiator  \cite{Kahn}, {$f_k$} denotes  the frequency of the $i$th subcarrier,
 ${\tau}$ is the signal propagation delay between the transmitter and receiver with
$\tau  = d/c$,
  ${{d}}$ is the distance between the transmitter and receiver, and
 $c$ is the speed of light, {$k = 0,...,2N-1$}.
The generalized Lambertian radiator ${g _{\rm{L}}}$
 can be expressed as
\begin{align}
{g _{\rm{L}}} =
\left\{ {\begin{array}{*{20}{c}}
   {\frac{{\left( {m + 1} \right){A_{\rm{r}}}\cos \left( \varphi  \right)}}{{2\pi {{{d}}^2}}}{{\cos }^m}\left( \theta  \right)T\left( \varphi  \right)G\left( \varphi  \right)}  \\
   0  \\
\end{array}} \right.\begin{array}{*{20}{c}}
   {0 \le \varphi  \le \Psi  },  \\
   {\rm{otherwise} },  \\
\end{array}
\end{align}
where
$m$ is the order of Lambertian emission, i.e., $m =  - \ln 2/\ln \left( {\cos {\Phi _{1/2}}} \right)$, ${{\Phi _{1/2}}}$ is the half power angle;
${A_{\rm{r}}}$ is the effective detector area of the PD receiver;
$\varphi$  and $\theta$ are, respectively, the incidence and irradiance angle from the LED to the PD;
${T\left( \varphi  \right)}$ and ${G\left( \varphi  \right)}$ are the optical filter gain and the concentrator gain of the receiver, respectively;
$\Psi $ represents the field-of-view (FOV) of the receiver.

On the other hand,
the gain of the  diffuse links {${H_{{\rm{D}},k}}$}  is given by \cite{Jungnickel}
\begin{align}\label{channel_model_diffuse}
{H_{{\rm{D}},k}} = \frac{{{\eta _D}}}{{1 + j2\pi \tau f_k}},
\end{align}
where ${\eta _{{\rm{D}}}}$  is the power efficiency of the diffuse signal {and $\tau$ is the exponential decay time. The time-domian diffuse channel gain $h_{{\rm{D}}}$  is given  as ${h_{{\rm{D}}}}\left( t \right) = \frac{{{\eta _D}}}{\tau }{e^{ - {t / \tau }}}\varepsilon \left( t \right)$, where $\varepsilon \left( t \right)$ is the unit step function.}

Thus, the time domain channel response can be given as $h\left( t \right) = {g_{\rm{L}}}\delta \left( t \right) + {h_{{\rm{D}}}}\left( {t - \Delta T} \right)$, where $\delta\left(t\right)$ is the Dirac function, and $\Delta T$ describes the delay between the LOS signal and   the diffuse signal. Besides, the relationship between the time domain channel response and the corresponding subcarriers channel gains  can be described as
$H\left( f_k \right) = \int_{ - \infty }^\infty  {h\left( t \right){e^{ - j2\pi f_k t}}\,\mathrm{d}t}$.

\subsection{Performance Metrics}
In practice, the  signals are transmitted from an LED through an
optical channel. At the receiver, it performs FFT  to obtain the frequency-domain modulated information.
However, due to the zero clipping, the amplitude of the frequency domain signal at the receiver is half of  that at the transmitter \cite{Jing}.
 Let $Y_{2i-1}$ {denotes} the signals received in the frequency-domain
   at the $\left( {2i - 1} \right)$th subcarrier, which   is given by
\begin{align}
{Y_{2i - 1}} = \frac{1}{2}{H_{2i - 1}}\sqrt {{p _{2i - 1}}} {X_{2i - 1}} + {Z_{2i - 1}},
\end{align}
where the coefficient $\frac{1}{2}$ exists since only   half subcarriers are adopted to transmit information,
 ${Z_{2i - 1}}$ is the additive white Gaussian noise (AWGN)  with  zero-mean, i.e., ${Z_{2i - 1}} \sim \mathcal{CN}\left( {0,{W}{\sigma ^2}} \right)$,
{$i = 1,...,N/2$}, and $\sigma^2$ represents the noise power spectral density, ${W}$ represents the bandwidth of each subcarrier.

Let ${{R_{2i - 1}}\left( {{\left\{ {{p _{2i - 1}}} \right\}_{i=1}^{N/2}}} \right)}$ and ${{ R}_{{\rm{ACO}}}}$  denote  the   rate of the $\left( {2i - 1} \right)$th subcarrier and the  total rate of  the ACO-OFDM system, respectively, which are    given by
  \begin{subequations}  \label{channel_capacity_ACO}
 \begin{align}
  &{R_{2i - 1}}\left( {{\left\{ {{p _{2i - 1}}} \right\}_{i=1}^{N/2}}} \right) = I\left( {{X_{2i - 1}};{Y_{2i - 1}}} \right), \hfill \\
  &{R_{{\text{ACO}}}} =   \sum\limits_{i = 1}^{N/2} {{R_{2i - 1}}\left( {{\left\{ {{p _{2i - 1}}} \right\}_{i=1}^{N/2}}} \right)},
\end{align}
\end{subequations}
respectively. Then, the SE of the ACO-OFDM VLC system
is defined as  the ratio of achievable
data rate to the total bandwidth,
which can be expressed as
\begin{align}   \label{se_band}
{\rm{SE}}\left( {{\left\{ {{p _{2i - 1}}} \right\}_{i=1}^{N/2}}} \right) = \frac{{\sum\limits_{i = 1}^{N/2} {{R_{2i - 1}}\left( {{\left\{ {{p _{2i - 1}}} \right\}_{i=1}^{N/2}}} \right)} }}{2NW} ,
\end{align}
where $W$ denotes the bandwidth of  each subcarrier.
At the same time, the EE of the ACO-OFDM VLC system is defined as the ratio of the   capacity
 to the total power consumption, which can be expressed as
\begin{align}   \label{ee_single}
{\rm{EE}}\left( {{\left\{ {{p _{2i - 1}}} \right\}_{i=1}^{N/2}}} \right) = \frac{{\sum\limits_{i = 1}^{N/2} {{R_{2i - 1}}\left( {{\left\{ {{p _{2i - 1}}} \right\}_{i=1}^{N/2}}} \right)} }}{{2\sum\limits_{i = 1}^{N/2} {{p _{2i - 1}}}}  + {P_c}},
\end{align}
where
${2\sum_{i = 1}^{N/2} {{p _{2i - 1}}}}$ represents the total electrical power consumption of all the  subcarriers and
${P_c}$ denotes the total circuit power consumption  of the whole system.

\section{Spectral efficiency of ACO-OFDM}

In this section, we aim to maximize the SE of  the ACO-OFDM system  under the electrical transmit  power constraint and taking into account a practical average optical power constraint. The considered problem can be mathematically  formulated as follows:
\begin{subequations} \label{aco_form}
\begin{align}
\mathop {{\text{maximize}}}\limits_{{{\left\{ {{p _{2i - 1}}} \right\}_{i=1}^{N/2}}}} &\quad \frac{{\sum\limits_{i = 1}^{N/2} {{R_{2i - 1}}\left( {{\left\{ {{p _{2i - 1}}} \right\}_{i=1}^{N/2}}} \right)} }}{2NW}  \label{PA_formu}\\
{\rm{s}}.{\rm{t}}.& \quad \mathbb{E}\left\{ {{{\hat x}_k}} \right\} \le  {P_o},\label{PA_con2_ACO}\\
&  \quad \sum\limits_{i = 1}^{N/2} {{p _{2i - 1}}}  \le P,\label{PA_con1}\\
&  \quad {p_{2i-1}} \ge 0, \quad i = 1,...,N/2.\label{PA_con3}
\end{align}
\end{subequations}

  In the following, we will investigate the SE maximization problem \eqref{aco_form} for the considered ACO-OFDM system
with  Gaussian distribution inputs and finite-alphabet inputs, respectively.

\subsection{Gaussian Distribution Inputs }
Assume that the input   ${X_{2i-1}}$ follows  independent complex Gaussian distribution, i.e., ${X_{2i-1}} \sim \mathcal{CN}\left( {0,1} \right)$.
 According to the IFFT  operation \eqref{IFFT_1},  the time domain signal  ${x_k}$  also follows Gaussian distribution, i.e., ${x_k} \sim \mathcal{N}\left( {0,{\frac{2}{{N}}\sum_{i = 1}^{N/2} {{p _{2i - 1}}} } } \right)$.
Furthermore, based on the relationship in  \eqref{zeroclip_ACO},  the average optical power  is given by\cite{Dimitrov2011ICC,armstrong2006power}
\begin{align}  \label{exp_ACO}
\mathbb{E}\left\{ {{{\hat x}_k}} \right\}& = \frac{1}{2}\mathbb{E}\left\{ {\left| {{x_k}} \right|} \right\}
= \frac{1}{2}\int_0^\infty  {x_k\frac{1}{\sqrt{2\pi}\sigma_s} {e^{ - \frac{{x_k^2}}{{2\sigma _s^2}}}}\,\mathrm{d}{x_k}}\nonumber\\
&= \sqrt {\frac{1}{{\pi N}}\sum\limits_{i = 1}^{N/2} {{p_{2i - 1}}} },
\end{align}
where $\sigma _s^2 = \frac{2}{N}\sum_{i = 1}^{N/2} {{p_{2i - 1}}} $.
Additionally, by substituting \eqref{exp_ACO} into \eqref{PA_con2_ACO},
the average optical power constraint can be reformulated as
 \begin{align}\label{optical_Ga}
 \sum\limits_{i = 1}^{N/2} {{p _{2i - 1}}}  \le {{N}\pi P_o^2}.
 \end{align}

According to the Shannon theorem \cite{CoverThormas}, the achievable rate of Gaussian distribution inputs
$R_\text{G}\left( {{{{p _{2i - 1}}}}} \right)$
   is given by
\begin{align} \label{Gaussian_ACO}
R_\text{G}\left( {{{{p _{2i - 1}}}}} \right) = {W}{\log _2}\left( {1 + \frac{{{p _{2i - 1}}{{\left| {{H_{2i - 1}}} \right|}^2}}}{{4{\sigma ^2}{W}}}} \right).
\end{align}

Then, the SE of the Gaussian distribution inputs ${\rm{S}}{{\rm{E}}_{\rm{G}}}\left( {{\left\{ {{p _{2i - 1}}} \right\}_{i=1}^{N/2}}} \right)$
can be expressed as
 \begin{align}    \label{se_Gau_1121}
 {\rm{S}}{{\rm{E}}_{\rm{G}}}\left( {{\left\{ {{p _{2i - 1}}} \right\}_{i=1}^{N/2}}}\right)= \frac{{\sum\limits_{i = 1}^{N/2} {{{\log }_2}\left( {1 + \frac{{{p _{2i - 1}}{{\left| {{H_{2i - 1}}} \right|}^2}}}{{4{\sigma ^2}{W}}}} \right)} }}{2N}.
 \end{align}

Thus, the SE maximization problem with the Gaussian distribution inputs can be rewritten as
\begin{subequations}\label{aco_form_G}
\begin{align}
\mathop {{\text{maximize}}
 }\limits_{{{\left\{ {{p _{2i - 1}}} \right\}_{i=1}^{N/2}}}} &\quad  {\rm{S}}{{\rm{E}}_{\rm{G}}}\left( {{\left\{ {{p _{2i - 1}}} \right\}_{i=1}^{N/2}}} \right)    \\
{\rm{s}}.{\rm{t}}.&\quad \sum\limits_{i = 1}^{N/2} {{p _{2i - 1}}}  \le \min \left\{ {P,{{{N}\pi P_o^2}}} \right\} , \label{simpaco_con1}\\
&\quad {p_{2i - 1}} \ge 0,\quad i = 1,...,N/2.\label{simpaco_con2}
\end{align}
\end{subequations}

Problem \eqref{aco_form_G} is a  convex optimization problem and
satisfies the Slater's constraint qualification \cite{Boyd}.
The problem in \eqref{aco_form_G} can be solved by applying classical convex optimization approaches.
To this end, we first need the
Lagrangian function of \eqref{aco_form_G}, which is given as
\begin{align}
{\cal{L}_{\rm{G}}} &= \frac{1}{2N}{\sum\limits_{i = 1}^{N/2} {{{\log }_2}\left( {1 + \frac{{{p_{2i - 1}}{{\left| {{H_{2i - 1}}} \right|}^2}}}{{4{\sigma ^2}{W}}}} \right)} } \nonumber\\
&~~~~~~~~~~~~- \mu \left( {\sum\limits_{i = 1}^{N/2} {{p_{2i - 1}}}  - \min \left\{ {P,{N}\pi P_o^2} \right\}} \right),
\end{align}
where  $\mu  \ge 0$  is the  Lagrange
multiplier   associated with  constraint \eqref{simpaco_con1}.
By setting the differential function to 0, i.e., $\frac{{\partial {\cal{L}_{\rm{G}}}}}{{\partial {p_{2i - 1}}}} = 0$,
the optimal ${p _{2i - 1}} $
is given by
\begin{align} \label{water_filling_p}
{p_{2i - 1}} = {\left[ {\frac{1}{{ 2N \mu \ln 2 }} - \frac{{4{\sigma ^2}{W}}}{{{{\left| {{H_{2i - 1}}} \right|}^2}}}} \right]^ + }.
\end{align}

 In fact, \eqref{water_filling_p} is known as the classical water-filling solution and
  the optimal $\mu$ can be found by the conventional gradient method or the  epsilon  method \cite{Boyd,Hamming1986}.

\subsection{Finite-alphabet Inputs}
In practice, typical inputs are always based on discrete
signaling constellations,   such as $M$-PSK or $M$-QAM,  rather than   the ideal Gaussian signals. In this section, we assume
   that  the inputs are drawn from   discrete
constellations set $\left\{ {{X_{2i - 1,k}}} \right\}_{k = 1}^M$
 with   cardinality $M$, where  $X_{2i-1,k}$ is a constellation point of the $\left({2i-1}\right)$th subcarrier.
   The achievable rate
   $R_\text{F}\left( {{{{p _{2i - 1}}}}} \right)$
 is given by \cite{Globally}
\begin{align}  \label{mutual_inf_ACO}
 R_\text{F}\left( {{{{p _{2i - 1}}}}} \right) & = {{I}_{2i-1}}\left({X_{2i - 1}}; {Y_{2i - 1}}  \right) \\
    & = {W}\left( {{{\log }_2}M - \frac{1}{{\ln 2}}} \right)\nonumber\\
    &~~~~~ - \sum\limits_{n = 1}^M {\frac{{{W}}}{M}{\mathbb{E}_Z}\left\{ {{{\log }_2}\sum\limits_{k = 1}^M {\exp \left( { - {d_{n,k}}} \right)} } \right\}},  \label{mutual_info_new}
 \end{align}
 where
 ${{I}_{2i-1}}\left({X_{2i - 1}}; {Y_{2i - 1}}  \right)$ is the achievable mutual information over the $\left({2i-1}\right)$th channel,
{${d_{n,k}} = \frac{1}{{{\sigma ^2}}{W}}{{{{\left| {\frac{1}{2}{H_{2i - 1}}\sqrt {{p _{2i - 1}}} \left( {{X_{2i-1,n}} - {X_{2i-1,k}}} \right) + {Z_{2i - 1}}} \right|}^2} } }$}
is a measure of the difference between input constellation points
$X_{2i-1,n}$ and $X_{2i-1,k}$,
${\mathbb{E}_Z}\left\{  \cdot  \right\}$ is the expectation of the noise ${Z_{2i - 1}}$.
 Note that ${{R_{{\rm{F}}}}\left( {{{{p _{2i - 1}}}}} \right)}$ is
 a concave function with respect to the power allocation ${p _{2i - 1}}$ \cite{Globally,Rajashekar}.

According to \eqref{zeroclip_ACO},
the average optical power of transmitted signals
is given as\cite{Dimitrov2011ICC,armstrong2006power}
 \begin{align}\label{ACO_aver}
 \mathbb{E}\left\{ {{\hat x}_k} \right\} & =   \frac{1}{2} \mathbb{E}\left\{ {\left| {{x_k}} \right|} \right\}\nonumber\\
  &\le   \frac{1}{{2\sqrt {2N} }}\sum\limits_{i = 0}^{2N - 1} {\mathbb{E}\left\{ {\left| {\sqrt {{p _i}} {X_i}\exp \left( {j\frac{{\pi ki}}{N}} \right)} \right|} \right\}} \nonumber\\
  &  =  \frac{1}{{2\sqrt {2N} }}\sum\limits_{i = 0}^{2N-1} {\sqrt {{p _i}} \mathbb{E} \left\{ {\left| {{X_i}} \right|} \right\}},
 \end{align}
 where the inequality   holds due to $\left| {\sum_i {{a_i}} } \right| \le \sum_i {\left| {{a_i}} \right|}$,
the value of  $\mathbb{E}\left\{ {\left| {{X_i}} \right|} \right\}$   depends on the specific modulation schemes{, i.e., \eqref{mutual_info_new} and \eqref{ACO_aver} can be apply on another distribution-known  discrete signaling constellations, such as OOK, DPSK, higher order PSK and QAM, and non-uniform discrete inputs}. Furthermore, substituting   \eqref{ACO_aver}  into \eqref{dimming_aco}, the   average optical power constraint   is given  as
\begin{align}\label{dimming_c_1}
\frac{1}{\sqrt{2N}}\sum\limits_{i = 1}^{N/2}
 {\sqrt {{p _{2i - 1}}} \mathbb{E}\left\{ {\left| {{X_{2i - 1}}} \right|} \right\}}  \le  {P_o}.
\end{align}

Based on the  inequality ${\left( {\sum_{i = 1}^n {{a_i}} } \right)^2} \le n\left( {\sum_{i = 1}^n {a_i^2} } \right)$ \cite{Chrystal}, where $ {a_i} \ge 0 $,
the average optical power constraint \eqref{dimming_c_1}
 can be restricted as
\begin{align} \label{MI_con3_jian}
\sum\limits_{i = 1}^{N/2} {{p _{2i - 1}}} \le \frac{{4P_o^2}}{{{\mathbb{E}^2}\left\{ {\left| {{X_{2i - 1}}} \right|} \right\}}}.
\end{align}

In other words, \eqref{MI_con3_jian} is known as a safe approximation of \eqref{dimming_c_1}  because the left side
  of \eqref{dimming_c_1} is replaced by its upper bound.
 After adopting the optimization algorithm, the optical power consumption would not exceed $P_o$.
 Then, the SE of finite-alphabet inputs ${\rm{S}}{{\rm{E}}_{\rm{F}}}\left(  {{\left\{ {{p _{2i - 1}}} \right\}_{i=1}^{N/2}}} \right)$
can be expressed as
\begin{align}
 {\rm{S}}{{\rm{E}}_{\rm{F}}}\left({{\left\{ {{p _{2i - 1}}} \right\}_{i=1}^{N/2}}} \right) = \frac{{\sum\limits_{i = 1}^{N/2} {{{R_{{\rm{F}}}}\left( {{{{p _{2i - 1}}}}} \right)}} }}{2NW}.
\end{align}

Thus, the  optimal power allocation  problem \eqref{aco_form}    can be reformulated  as   the
constellation-constrained mutual information maximization problem  which
 can be expressed as follows
\begin{subequations}\label{aco_finite}
\begin{align}
\mathop {{\text{maximize}}
 }\limits_{{{\left\{ {{p _{2i - 1}}} \right\}_{i=1}^{N/2}}}} &\quad {\rm{S}}{{\rm{E}}_{\rm{F}}}\left( {{\left\{ {{p _{2i - 1}}} \right\}_{i=1}^{N/2}}}\right)   \label{aco_finite_a}\\
{\rm{s}}{\rm{.t}}{\rm{. }} &\quad \sum\limits_{i = 1}^{N/2} {{p _{2i - 1}}}  \le \varepsilon,  \label{power_constr}\\
&\quad p_{2i - 1} \ge 0{\rm{ }},\quad i = 1,...,N/2,
\end{align}
\end{subequations}
where $\varepsilon  \buildrel \Delta \over = \min \left\{ {P,\frac{{4P_o^2}}{{{\mathbb{E}^2}\left\{ {\left| {{X_{2i - 1}}} \right|} \right\}}}} \right\}$.

The lack of a closed-form expression
 for the objective function    \eqref{aco_finite_a}   complicates its solution development.
To address this difficulty, we aim to derive the optimal power allocation scheme for problem \eqref{aco_form}    by exploiting the relationship between   the mutual information    and  MMSE  \cite{Shamai}.

 To this end, we first derive
 the  {equivalent} Lagrangian function of   problem \eqref{aco_finite} which is given by
\begin{align}
\mathcal{L}_{\rm{F}} =&   - \sum\limits_{i = 1}^{N/2} {R_{{\rm{F}}}}\left( {{{{p _{2i - 1}}}}} \right)  + \lambda \left( {\sum\limits_{i = 1}^{N/2} {{p _{2i - 1}}}  - \varepsilon } \right),
\end{align}
where $\lambda  \ge 0$
  is the   Lagrange multiplier corresponding to constraint \eqref{power_constr}.

Furthermore, the KKT conditions of problem \eqref{aco_finite} can be expressed as
\begin{subequations}
\begin{align}
&\quad - \frac{{\partial {R_{{\rm{F}}}}\left( {{{{p _{2i - 1}}}}} \right)}}{{\partial {p _{2i - 1}}}} + \lambda   = 0, \label{acoKKT_1}\\
&\quad \lambda \left( {\sum\limits_{i = 1}^{N/2}  {{p _{2i - 1}}}  - \varepsilon } \right) = 0,\\
&\quad \sum\limits_{i = 1}^{N/2}  {{p _{2i - 1}}}  - \varepsilon \le 0,\\
&\quad \lambda  \ge 0,\quad p_{2i - 1} \ge 0, \quad i = 1,...,N/2.
\end{align}
\end{subequations}

According to \cite{Shamai}, the  relationship between the  mutual information  and the  MMSE of the $(2i-1)$th subcarrier  is given by
\begin{align} \label{MMSE_MI_aco}
\frac{\partial }{{\partial {\rm{SNR}}}}{I_{2i - 1}}\left( {{X_{2i - 1}};{Y_{2i - 1}}} \right) = {\rm{MMS}}{{\rm{E}}_{2i{\rm{ - }}1}}\left( {{\rm{SNR}}} \right),
\end{align}
where
${\rm{MMS}}{{\rm{E}}_{2i - 1}}\left( {\rm{SNR}}  \right) =
\mathbb{E}\left\{ {{{\left| {{X_{2i - 1}} - {{\hat X}_{2i - 1}}} \right|}^2}} \right\}$
is  the  MMSE  of  ${X_{2i - 1}}$,
and ${{\hat X}_{2i - 1}}$ is conditional expectation of  ${X_{2i - 1}}$, i.e.,
\begin{align}
 {{\hat X}_{2i - 1}}
 & = \mathbb{E}\left\{ {{X_{2i - 1}}\left| {{Y_{2i - 1}} = \frac{1}{2}{H_{2i - 1}}\sqrt {{p _{2i - 1}}}  {X_{2i - 1}} + {Z_{2i - 1}}} \right.} \right\}.
\end{align}
Combining \eqref{mutual_inf_ACO} and \eqref{MMSE_MI_aco},
the differential function of ${R_{{\rm{F}}}}\left( {{{{p _{2i - 1}}}}} \right)$ can be written as
\begin{align}\label{MMSE_MI_aco_2}
\frac{{\partial {R_{{\rm{F}}}}\left( {{{{p _{2i - 1}}}}} \right)}}{{\partial {p _{2i - 1}}}} &= \frac{{{{\left| {{H_{2i - 1}}} \right|}^2}}}{{4{\sigma ^2}{W}}}{\rm{MMS}}{{\rm{E}}_{2i - 1}}\left( {\frac{{{{\left| {{H_{2i - 1}}} \right|}^2}}}{{4{\sigma ^2}{W}}}{p_{2i - 1}}} \right).
\end{align}

By substituting \eqref{MMSE_MI_aco_2} into \eqref{acoKKT_1},
 we have
 \begin{align} \label{MMSE_p}
 \frac{{{{\left| {{H_{2i - 1}}} \right|}^2}}}{{4{\sigma ^2}{W}}}{\rm{MMS}}{{\rm{E}}_{2i - 1}}\left( {\frac{{{{\left| {{H_{2i - 1}}} \right|}^2}}}{{4{\sigma ^2}{W}}}{p_{2i - 1}}} \right) = \lambda.
 \end{align}

Then, solving \eqref{MMSE_p} for the  power allocation $p _{2i - 1}$ yields
\begin{align}   \label{alpha_aco}
p _{2i - 1} = \frac{{4{\sigma ^2}{W}}}{{{{\left| {{H_{2i - 1}}} \right|}^2}}}{\rm{MMSE}}_{2i - 1}^{ - 1}\left( {\frac{{ 4 {\sigma ^2}{W}}}{{{{\left| {{H_{2i - 1}}} \right|}^2}}}\lambda} \right),
\end{align}
where ${\rm{MMSE}}_{2i - 1}^{ - 1}\left(  \cdot  \right)$ is
  the inverse function of ${\rm{MMS}}{{\rm{E}}_{2i - 1}}\left(  \cdot  \right)$
with the domain  $\left[ {0,1} \right]$ and ${\rm{MMSE}}_{2i - 1}^{ - 1}\left( 1 \right) = 0$ \cite{Lozano}.

 Therefore, for the considered ACO-OFDM system, the optimal power allocation scheme of \eqref{aco_finite} is given by
\begin{align} \label{ai}
p _{2i - 1}^* =
\begin{cases}
    {\frac{{4{\sigma ^2}{W}}}{{{{\left| {{H_{2i - 1}}} \right|}^2}}}{\rm{MMSE}}_{2i - 1}^{ - 1}\left( {\frac{{ 4 {\sigma ^2}{W}}}{{{{\left| {{H_{2i - 1}}} \right|}^2}}}\lambda} \right)},{~0 < \lambda  \leq \frac{{{{\left| {{H_{2i - 1}}} \right|}^2}}}{{4{\sigma ^2}{W}}};}\\
    ~~~~~~~~~~~~~~~~~~~~~~~{0},{~~~~~~~~~~~~~~~{\rm{otherwise}.}}
\end{cases}
\end{align}

The dual variable  $\lambda$ in  \eqref{ai}  is the solution of the following equation
\begin{align} \label{lambda_1}
\sum\limits_{i = 1}^{N/2} \frac{{4{\sigma ^2}{W}}}{{{{\left| {{H_{2i - 1}}} \right|}^2}}}{\rm{MMSE}}_{2i - 1}^{ - 1}\left( {\frac{{ 4 {\sigma ^2}{W}}}{{{{\left| {{H_{2i - 1}}} \right|}^2}}}\lambda} \right)  = \varepsilon,
\end{align}
which can be obtained  by  a simple bisection method as listed in   Algorithm \ref{bisection_method_alg}.
\begin{algorithm}[htbp]
    \caption{Bisection Method}
    \label{bisection_method_alg}
    \begin{algorithmic}[1]
        \Require  Given $\lambda  \in \left[ {0,\hat \lambda } \right]$, $\delta  > 0$, and initialize  ${\lambda _{\min }} = 0,{\lambda _{\max }} = \hat \lambda $,
        where $\delta$ denotes termination parameter, and  $\hat{\lambda}$ represents an upper bound of $\lambda$;
        \While{${\lambda _{\max }} - {\lambda _{\min }} \ge \delta $}
            \State Set $\lambda  = \left( {{1 \mathord{\left/ {\vphantom {1 2}} \right. \kern-\nulldelimiterspace} 2}} \right)\left( {{\lambda _{\min }} + {\lambda _{\max }}} \right)$;
          \State If $\lambda  < {{{{\left| {{H_{2i - 1}}} \right|}^2}}}/{4{\sigma ^2}{W}}$,
             substitute $\lambda$ to  obtain $p _{2i - 1}^* = \frac{4{\sigma ^2}{W}}{{\left| {{H_{2i - 1}}} \right|}^2}{\rm{MMSE}}_{2i - 1}^{ - 1}\left( {\frac{ 4 {\sigma ^2}{W}}{{\left| {{H_{2i - 1}}} \right|}^2}\lambda} \right)$;
             otherwise $p _{2i - 1}^* = 0$;
          \State If $\sum\limits_{i = 1}^{N/2} {p _{2i - 1}^*}  \le \varepsilon $,
             set ${\lambda _{\max }} \leftarrow \lambda $; otherwise ${\lambda _{\min }} \leftarrow \lambda $;
        \EndWhile
        \Ensure $p _{2i - 1}^*$;
    \end{algorithmic}
\end{algorithm}

In the following, we explain  the  differences  between the    water-filling power allocation  \eqref{water_filling_p} with  Gaussian distribution inputs and the   power allocation \eqref{ai}   with finite-alphabet inputs.
To facilitate the presentation, let us introduce a function, ${G_{2i-1}}\left( \lambda  \right)$, as follows
\begin{align}\label{G_F}
{G_{2i-1}}\left( \lambda  \right) =
\begin{cases}
    \frac{{{{\left| {{H_{2i - 1}}} \right|}^2}}}{{4{\sigma ^2}{W}\lambda }} - {\rm{MMS}}{{\rm{E}}_{2i - 1}^{{\rm{ - 1}}}}\left( {\frac{{4{\sigma ^2}{W}\lambda }}{{{{\left| {{H_{2i - 1}}} \right|}^2}}}} \right),\\
    {~~~~~~~~~~~~~~~~~~~~~~0 <  \lambda  \leq  \frac{{{{\left| {{H_{2i - 1}}} \right|}^2}}}
        {{4{\sigma ^2}{W}}};}\\
    {~~~~~~~~~~~~~~~1},{~~~~{\rm{otherwise}.}}
\end{cases}
%\left\{ {\begin{array}{*{20}{c}}
%   {\frac{{{{\left| {{H_{2i - 1}}} \right|}^2}}}{{4{\sigma ^2}{W}\lambda }} - {\rm{MMS}}{{\rm{E}}_{2i - 1}^{{\rm{ - 1}}}}\left( {\frac{{4{\sigma ^2}{W}\lambda }}{{{{\left| {{H_{2i - 1}}} \right|}^2}}}} \right),}
%   & {~0 <  \lambda  \leq  \frac{{{{\left| {{H_{2i - 1}}} \right|}^2}}}
%{{4{\sigma ^2}{W}}};}  \\
%   {1}, & {{\rm{otherwise}.}}  \\
%\end{array}} \right.
\end{align}
 Then, the  power allocation  ${p_{2i - 1}}$  can be represented as
\begin{align}\label{P_unif}
{p_{2i - 1}} = \frac{1}{\lambda } - \frac{{4{\sigma ^2}{W}}}{{{{\left| {{H_{2i - 1}}} \right|}^2}}}{G_{2i - 1}}\left( \lambda  \right).
\end{align}

\begin{figure}[htbp]
    \centering
    \includegraphics[width =0.5\textwidth]{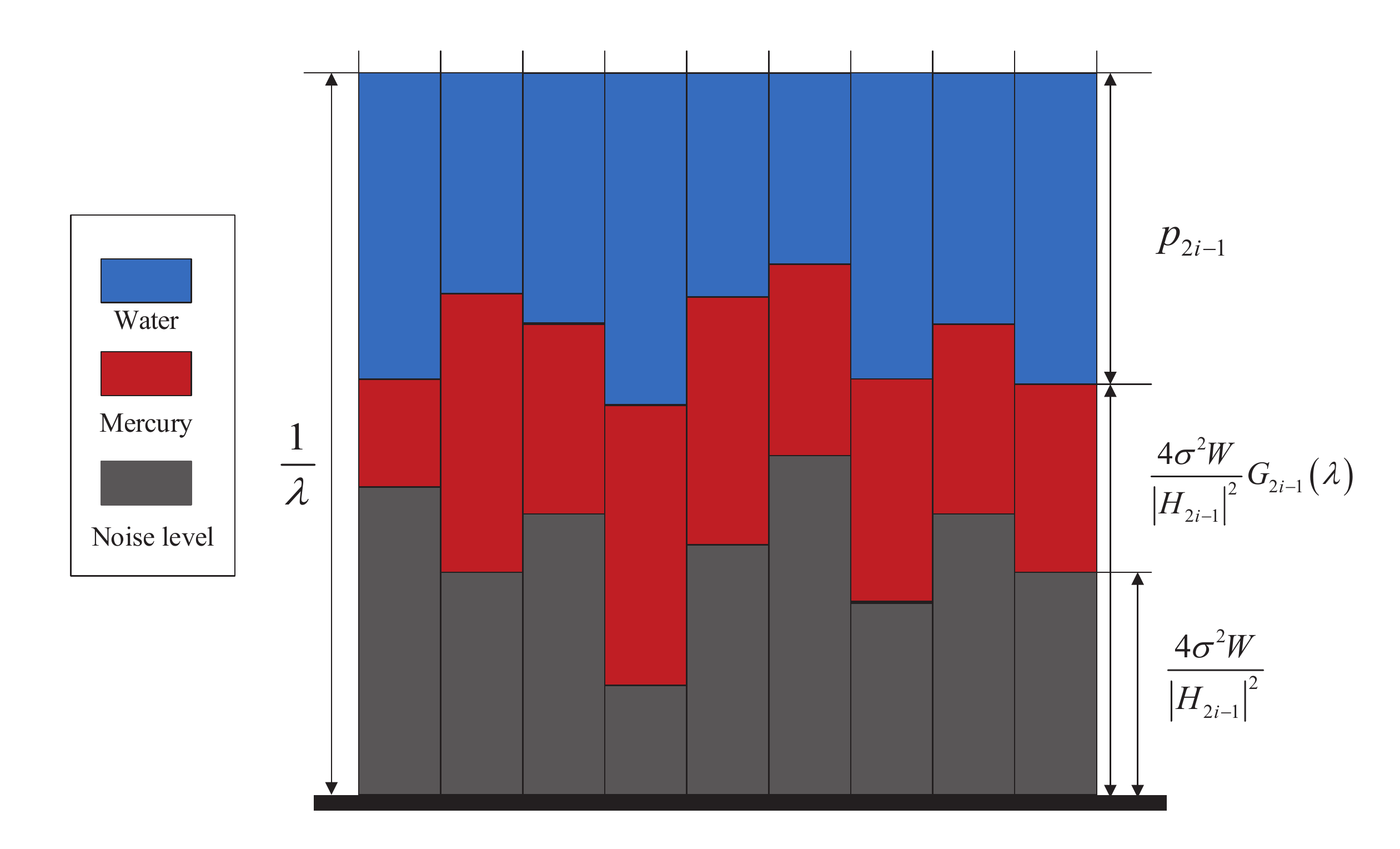}
    \caption{~ Mercury water-filling scheme.}
    \label{fig_mercury_water}
\end{figure}

Note that  if ${G_{2i-1}}\left( \lambda  \right) = 1$, the  power allocation \eqref{P_unif}
is equivalent to the water-filling power allocation \eqref{water_filling_p} with   the Gaussian distribution inputs; otherwise, the  power allocation \eqref{P_unif}
is the   power allocation   \eqref{ai}   with finite-alphabet inputs\footnote{
The ACO-OFDM with finite-alphabet inputs converges weakly to that with Gaussian distribution inputs
as the number of
subcarriers $N$ is sufficiently large \cite{Wei_TIT_2010}.}.
Furthermore, based on the function  ${G_{2i-1}}\left( \lambda  \right)$,   the  power allocation   \eqref{P_unif} can be interpreted as the mercury-water-filling scheme \cite{Lozano}. More specifically,
the power allocation \eqref{P_unif} is illustrated in Fig.~\ref{fig_mercury_water}.
For the $\left({2i-1}\right)$th subcarrier with noise level $\frac{{4{\sigma ^2}{W}}}
{{{{\left| {{H_{2i - 1}}} \right|}^2}}}$, we first fill mercury  to the height $\frac{{4{\sigma ^2}{W}}}{{{{\left| {{H_{2i - 1}}} \right|}^2}}}{G_{2i - 1}}\left( \lambda  \right)$, then, pouring water (power) to the  height $\frac{1}
{\lambda }$.  Note that both the noise level and the mercury in the mercury-water-filling scheme form the
 bottom   level of  the water-filling   scheme.

\subsection{Lower Bound of Mutual Information}

For finite-alphabet  inputs, the optimal power allocation scheme \eqref{ai} involves
the calculation of integrals of the MMSE ranged from $ - \infty$
 to $ + \infty $, which can only be obtained by {Monte Carlo method}
 and numerical integral methods at the expense of high computational  complexity \cite{Globally,Lozano}.
To strike a balance between complexity and performance, we further
develop a low complexity power allocation scheme.

As mentioned above,  the expression of  mutual information of ACO-OFDM is given in \eqref{mutual_info_new}.
The  upper bound   of the expectation term in \eqref{mutual_info_new}  is given as \cite{XiaoLu}
 \begin{subequations}
\begin{align}
&{{\mathbb{E}}_Z}\left\{ {\log _2}\sum\limits_{k = 1}^M { \exp \left( { - {d_{nk}}} \right)}\right\}
\le {\log _2}\sum\limits_{k = 1}^M { {{\mathbb{E}}_{Z_{2i-1}}} \left\{ \exp \left( { - {d_{nk}}} \right) \right\} } \label{jense}\\
& = {\rm{ lo}}{{\rm{g}}_{\rm{2}}}\sum\limits_{k = 1}^M { {\int_{Z_{2i-1}} {\frac{\exp \left( { - {d_{nk}}} \right)}{{ {\pi } \sigma^2{W} }}\exp \left( { - \frac{{{{\left| {{Z_{2i - 1}}} \right|}^2}}}{{{\sigma ^2{W}}}}} \right)} \,\mathrm{d}Z_{2i-1}} } \label{exp_z}\\
& =  {\log _2}\sum\limits_{k = 1}^M {\frac{1}{2}} \exp \left( { - \frac{{{{{\mathop{\rm Re}\nolimits} }^2}\left\{ {C_{2i - 1}} \right\} + {{{\mathop{\rm Im}\nolimits} }^2}\left\{ {C_{2i - 1}} \right\}}}{{2{\sigma ^2}{W}}}} \right)\\
& =- 1 + {\rm{lo}}{{\rm{g}}_{\rm{2}}}\sum\limits_{k = 1}^M { {\exp \left( { - \frac{{ {p _{2i - 1}}{{{{\left| {{H_{2i - 1}}} \right|}^2}}} {{\left| {{X_{2i-1,n}} - {X_{2i-1,k}}} \right|}^2}}}{{8{\sigma ^2}{W}}}} \right)} } \label{exp_upper},
\end{align}
\end{subequations}
where {${C_{2i - 1}} \buildrel \Delta \over = \frac{1}{2}{H_{2i - 1}}\sqrt {{p_{2i - 1}}} \left( {{X_{2i - 1,n}} - {X_{2i - 1,k}}} \right)$}, inequality   \eqref{jense}  is based on Jensen's inequality and \eqref{exp_z} is the expectation over $Z$.{Besides, the above derivation can also be used for another distribution-known  discrete signaling constellations, such as OOK, DPSK, and non-uniform discrete inputs.}
Note that the upper bound in \eqref{exp_upper} is a detertimintic value without involving integration
which can be adopted in the following for the development of computationally efficient resource allocation algorithm.
Let ${{{ R}_{\rm{L}}}\left( {{{{p _{2i - 1}}}}} \right)}$ represents the lower bound of mutual information
in the $\left( {{2i - 1}} \right)$th subcarrier. Thus, the lower bound of achievable rate of  the ACO-OFDM VLC system   with finite-alphabet inputs is given by
\begin{align}\label{total_bound_ACO}
 &{R_{\rm{L}}}\left( {{{{p _{2i - 1}}}}} \right) = {W}\left( {{\log }_2}M + 1 - \frac{1}{{\ln 2}}\right)-\nonumber\\
 & \sum\limits_{n = 1}^M {\frac{W}{M} {{\log }_2}\sum\limits_{k = 1}^M { {\exp \left( { - \frac{{ {p _{2i - 1}}{{{{\left| {{H_{2i - 1}}} \right|}^2}}} {{\left| {{X_{2i-1,n}} - {X_{2i-1,k}}} \right|}^2}}}{{8{\sigma ^2}{W}}}} \right)} } } .
\end{align}

Then, the SE of the lower bound of mutual information ${\rm{S}}{{\rm{E}}_{\rm{L}}}\left( {{\left\{ {{p _{2i - 1}}} \right\}_{i=1}^{N/2}}} \right)$
can be expressed as
\begin{align}  \label{lower_sub}
 {\rm{S}}{{\rm{E}}_{\rm{L}}}\left({{\left\{ {{p _{2i - 1}}} \right\}_{i=1}^{N/2}}}\right) = \frac{{\sum\limits_{i = 1}^{N/2} {{{R_{{\rm{L}}}}\left( {{{{p _{2i - 1}}}}} \right)}} }}{2NW}.
\end{align}

By substituting the   derived lower bound \eqref{total_bound_ACO}  into  the objective of
   problem of \eqref{lower_sub},
the SE maximization problem \eqref{aco_form} can be reformulated as follows
\begin{subequations}   \label{finite_lowerbound_ACO}
\begin{align}
\mathop {{\text{maximize}}
 }\limits_{{{\left\{ {{p _{2i - 1}}} \right\}_{i=1}^{N/2}}}} ~& {\rm{S}}{{\rm{E}}_{\rm{L}}}\left( {{\left\{ {{p _{2i - 1}}} \right\}_{i=1}^{N/2}}} \right)   \\
 {\rm{s}}{\rm{.t}}{\rm{.}}{\rm{    }}~& \sum\limits_{i = 1}^{N/2} {{p _{2i-1}}}  \le  \varepsilon,  \\
~&  {p _{2i - 1}} \ge 0, \quad   i = 1,...,N/2,
\end{align}
\end{subequations}
where $\varepsilon \buildrel \Delta \over = \min \left\{ {P,\frac{{4P_o^2}}{{{\mathbb{E}^2}\left\{ {\left| {{X_{2i - 1}}} \right|} \right\}}}} \right\}$.
This optimization problem \eqref{finite_lowerbound_ACO} is a standard convex problem, which can be
efficiently solved by the interior-point algorithm \cite{Boyd,Grantexample}. {It can also be solved by the mercury-water-filling scheme similar to the one adopted in Section III-B.}

\section{Energy Efficiency of ACO-OFDM}

In this section,
we investigate the design of optimal power allocation scheme to maximize   the EE of  the ACO-OFDM VLC system which is  subject
to   the QoS requirement,and both the electrical and optical power constraints.
Mathematically, we can formulate the EE  maximization problem of the  ACO-OFDM VLC system  as
 \begin{subequations} \label{EE_ori}
 \begin{align}
 \mathop {{\text{maximize}}
 }\limits_{{{\left\{ {{p _{2i - 1}}} \right\}_{i=1}^{N/2}}}} ~& \frac{{\sum\limits_{i = 1}^{N/2} {{R_{2i - 1}}\left({{\left\{ {{p _{2i - 1}}} \right\}_{i=1}^{N/2}}} \right)} }}{{2\sum\limits_{i = 1}^{N/2} {{p _{2i - 1}}}  + {P_c}}}  \label{ee_ori_conxx1}\\
 {\rm{s}}{\rm{.t}}{\rm{.}}{\rm{    }}~& \mathbb{E}\left\{ {{{\hat x}_k}} \right\} \le  {P_o}, \label{EE_ori_con0}\\
 & \sum\limits_{i = 1}^{N/2} {{p _{2i - 1}}}  \le P,  \label{EE_ori_con1} \\
 & {\sum\limits_{i = 1}^{N/2} {{R_{2i - 1}}\left( {{\left\{ {{p _{2i - 1}}} \right\}_{i=1}^{N/2}}} \right)} }  \ge r,  \label{EE_ori_con2} \\
 & {p _{2i - 1}} \ge 0, \quad i = 1,...,N/2,
\end{align}
\end{subequations}
where $r$ is the minimum achievable rate requirement.

Problem \eqref{EE_ori} is a general formulation of the EE in ACO-OFDM VLC systems.
In the following, we will present the optimal EE scheme of   problem \eqref{EE_ori}
with Gaussian distribution inputs and finite-alphabet inputs respectively.
Note that  both objective function \eqref{ee_ori_conxx1} and the constraint  \eqref{EE_ori_con2} of  the  EE maximization problem \eqref{EE_ori} are  different from  that of  the  SE maximization  problem \eqref{aco_form}.

\subsection{Gaussian Distribution Inputs}

With   Gaussian distribution inputs and  the achievable rate expression    \eqref{Gaussian_ACO},
the EE of Gaussian distribution ${\rm{E}}{{\rm{E}}_{\rm{G}}}\left( {{\left\{ {{p _{2i - 1}}} \right\}_{i=1}^{N/2}}} \right)$
can be expressed as
\begin{align}     \label{ee_Gau_1121}
{\rm{E}}{{\rm{E}}_{\rm{G}}}\left( {{\left\{ {{p _{2i - 1}}} \right\}_{i=1}^{N/2}}} \right) = \frac{{\sum\limits_{i = 1}^{N/2} {{W}{{\log }_2}\left( {1 + \frac{{{p _{2i - 1}}{{{{\left| {{H_{2i - 1}}} \right|}^2}}}}}{4{{\sigma ^2}}{W}}} \right)} }}{{2\sum\limits_{i = 1}^{N/2} {{p _{2i - 1}}}  + {P_c}}}.
\end{align}

  The     EE maximization problem with Gaussian distribution inputs of ACO-OFDM systems can be formulated as
 \begin{subequations} \label{EE_ACOOFDM}
 \begin{align}
 \mathop {{\text{maximize}}
 }\limits_{{{\left\{ {{p _{2i - 1}}} \right\}_{i=1}^{N/2}}}} ~& {\rm{E}}{{\rm{E}}_{\rm{G}}}\left( {{\left\{ {{p _{2i - 1}}} \right\}_{i=1}^{N/2}}} \right)  \label{EE_ACOOFDM_Gau}\\
 {\rm{s}}{\rm{.t}}{\rm{.}}{\rm{    }}~& \sum\limits_{i = 1}^{N/2} {{p _{2i - 1}}}  \le \min \left\{ {P,{{{N}\pi P_o^2}}} \right\},  \label{EE_ACOOFDM_con1}\\
 & {\sum\limits_{i = 1}^{N/2} {{W}{{\log }_2}\left( {1 + \frac{{{p _{2i - 1}}{{{{\left| {{H_{2i - 1}}} \right|}^2}}}}}{4{{\sigma ^2}}{W}}} \right)} }  \ge r,  \label{EE_ACOOFDM_con2}\\
 & {p _{2i - 1}} \ge 0, \quad i = 1,...,N/2. \label{EE_ACOOFDM_con3}
\end{align}
\end{subequations}

   For the  objective function \eqref{EE_ACOOFDM_Gau}, the   numerator is a differentiable concave function of the variable ${{p _{2i - 1}}}$,
while the denominator is an affine function of ${{p _{2i - 1}}}$.
Thus,
the   objective function \eqref{EE_ACOOFDM_Gau}   is a quasi-concave function of ${{p _{2i - 1}}}$.
 With a convex constraint set,
   problem  \eqref{EE_ACOOFDM} is a typical fractional problem\cite{Ng}, which is generally non-convex.
  To circumvent the non-convexity, Dinkelbach-type iterative algorithm \cite{Dinkelbach,Zappone,Crouzeix} can be adopted to trackle the problem
by converting the problem \eqref{EE_ACOOFDM} into a sequence of convex subproblems.
In particular, solving these convex subproblems iteratively
can eventually obtain the globally optimal solution of problem \eqref{EE_ACOOFDM}.
 Specifically, let $\Upsilon $ denote the feasible set defined  constraints of problem \eqref{EE_ACOOFDM} as follows
\begin{align}
\Upsilon =\left\{ p_{2i-1} \left|\eqref{EE_ACOOFDM_con1}, \eqref{EE_ACOOFDM_con2}, \eqref{EE_ACOOFDM_con3}, i = 1,...,N/2\right.\right\}.
\end{align}

Moreover, we introduce  a new function $f\left( {{\left\{ {{p _{2i - 1}}} \right\}_{i=1}^{N/2}}} \right)$ as follows
 \begin{align}
&f\left( {{\left\{ {{p _{2i - 1}}} \right\}_{i=1}^{N/2}}} \right) \triangleq\nonumber\\
& \sum\limits_{i = 1}^{N/2} { {W} {{\log }_2}\left( {1 + \frac{{{p _{2i - 1}}{{{\left| {{H_{2i - 1}}} \right|}^2}}}}{4{{\sigma ^2}{W}}}} \right)}- q \left( {2\sum\limits_{i = 1}^{N/2} {{p _{2i - 1}}}  + P_c} \right),
 \end{align}
where $q$ is a given non-negative parameter to be found iteratively.
Then, by calculating the roots of the equation
$f\left( {{\left\{ {{p _{2i - 1}}} \right\}_{i=1}^{N/2}}} \right) = 0$ in the set $\Upsilon $,  the optimal solution of problem \eqref{EE_ACOOFDM} can be obtained.

For a given $q$ in each iteration, the convex subproblem over ${{p_{2i-1}}}$ can be expressed as
\begin{subequations} \label{dinkelbach_ACO_Gaussian}
\begin{align}
\mathop {{\text{maximize}}
 }\limits_{{{\left\{ {{p _{2i - 1}}} \right\}_{i=1}^{N/2}}}}~& f\left( {{\left\{ {{p _{2i - 1}}} \right\}_{i=1}^{N/2}}} \right)\\
{\rm{s}}{\rm{.t}}{\rm{.   }} ~&{\forall{p _{2i - 1}} \in \Upsilon , i = 1,...,N/2}.
\end{align}
\end{subequations}

  Since the transformed problem is convex and satisfies the Slater's  constraint qualification,
 we apply the conventional optimization techniques by
 taking the partial derivative of function $f\left( {{\left\{ {{p _{2i - 1}}} \right\}_{i=1}^{N/2}}} \right)$ and setting it to zero, i.e.
 $\frac{{\partial f\left( {{\left\{ {{p _{2i - 1}}} \right\}_{i=1}^{N/2}}} \right)}}{{\partial {p _{2i - 1}}}} = 0,$
 which yeilds
 \begin{align}  \label{ee_Gau_alpha}
 {{\widetilde p }_{2i - 1}} = {\left[ {\frac{{W}}{{2q \ln 2}} - \frac{{4{\sigma ^2}{W}}}{{{{\left| {{H_{2i - 1}}} \right|}^2}}}} \right]^ + }.
 \end{align}

Then, by projecting ${{\widetilde p }_{2i - 1}}$ into the feasible region $\Upsilon$, we obtain the optimal power $p _{2i - 1}^*$ of problem \eqref{dinkelbach_ACO_Gaussian} as follows
 \begin{align}\label{ee_Gau_alpha_P}
    \left\{ p_{2i - 1}^* \right\}_{i = 1}^{N/2} &= \mathrm{Proj}_{\Upsilon }\left(\left\{ \widetilde{p}_{2i - 1} \right\}_{i = 1}^{N/2}\right)\nonumber\\
    & = \mathop {\arg \min }\limits_{\left\{ p_{2i - 1} \right\}_{i = 1}^{N/2}} \sum\limits_{i=1}^{N/2}\left\|\widetilde{p}_{2i - 1} - p_{2i-1}\right\|^2
\end{align}
where $\mathrm{Proj}_{\Upsilon }\left(\left\{ \widetilde{p}_{2i - 1} \right\}_{i = 1}^{N/2}\right) $ denotes
 the projection of     $\left\{ \widetilde{p}_{2i - 1} \right\}_{i = 1}^{N/2}$ into the subspace $\Upsilon $.

Note that if  $\left\{ \widetilde{p}_{2i - 1} \right\}_{i = 1}^{N/2}\in \Upsilon$,
the power allocation \eqref{ee_Gau_alpha_P} for the EE maximization is a generalization of the  power allocation   \eqref{water_filling_p} for the SE maximization.
In particular, for a small electrical transmit power budget, both SE and EE maximization show the same water-filling solution.
However, when the budget is sufficiently large, once the maximum EE is achieved,
the optimal EE power algorithm would clip the transmit power level, as can be observed in \eqref{ee_Gau_alpha_P}.

Finally, the EE maximization problem  of Gaussian distribution inputs can be solved by the Dinkelbach-type algorithm.
Under a finite number of iterations,
 the Dinkelbach-type algorithm is guaranteed to converge to the optimal solution of problem \eqref{EE_ACOOFDM}, e.g. \cite{Dinkelbach,Zappone,Crouzeix}.
 Algorithm \ref{dinkelbach_alg} shows the detail of  implementation.

%%%%%%%%%%%%%%%%%%%%%%%%%%%%%%%%%%%%%%%%%%%%%%%%%%%%%%%%%

\begin{figure*}[!b]
    \hrulefill
    \normalsize
    \setcounter{mytempeqncnt}{\value{equation}}
    % Eq. (53)
    \setcounter{equation}{52}
    \begin{align}     \label{EE_F_long}
        \mathrm{EE}_\mathrm{F}\left( {{\left\{ {{p _{2i - 1}}} \right\}_{i=1}^{N/2}}}\right)= \frac{ {\frac{NW}{2}\left(  {{\log }_2}M - \frac{1}{{\ln 2}} \right)- \sum\limits_{i = 1}^{N/2}\sum\limits_{n = 1}^M {\frac{W}{M}{\mathbb{E}_Z}\left\{ {{{\log }_2}\sum\limits_{k = 1}^M  \exp \left( { - \frac{{{{\left| {\frac{1}{2}{H_{2i - 1}}\sqrt {{p_{2i - 1}}} \left( {{X_{2i - 1,n}} - {X_{2i - 1,k}}} \right) + {Z_{2i - 1}}} \right|}^2}}}{{{\sigma ^2}{W}}}} \right)} \right\}}  }  }{2\sum\limits_{i = 1}^{N/2} p _{2i - 1}  + P_c}.
    \end{align}

    % Eq. (55)
    \setcounter{equation}{54}
    \begin{align}     \label{EE_ACO}
        {\rm{E}}{{\rm{E}}_{\rm{L}}}\left( {{\left\{ {{p _{2i - 1}}} \right\}_{i=1}^{N/2}}} \right)= \frac{{\frac{N{W}}{2}\left( { {{\log }_2}M + 1 - \frac{1}{{\ln 2}}} \right) - \sum\limits_{i = 1}^{N/2} {\sum\limits_{n = 1}^M {\frac{{W}}{M} {{\log }_2}\sum\limits_{k = 1}^M { \exp \left( { - \frac{{{{\left| {\frac{1}{2}{H_{2i - 1}}\sqrt {{p_{2i - 1}}} \left( {{X_{2i - 1,n}} - {X_{2i - 1,k}}} \right)} \right|}^2}}}{{2{\sigma ^2}{W}}}} \right)} } } }}{{2\sum\limits_{i = 1}^{N/2} {{p _{2i - 1}}}  + {P_c}}}.
    \end{align}
\end{figure*}
\setcounter{equation}{\value{mytempeqncnt}}
%%%%%%%%%%%%%%%%%%%%%%%%%%%%%%%%%%%%%%%%%%%%%%%%%%%%%%%%%

\begin{algorithm}[htbp]
    \caption{Dinkelbach-type Algorithm}
    \label{dinkelbach_alg}
    \begin{algorithmic}[1]
        \Require  Given $\delta  \to 0,n = 0,p _{2i - 1}^* > 0,{q ^{\left( n \right)}} = 0$;
        \While{$\left| {{q ^{\left( n \right)}} - {q ^{\left( {n + 1} \right)}}} \right| \le \delta $}
            \State Compute the optimal solution ${\left\{ p_{2i - 1}^* \right\}_{i = 1}^{N/2}} $;
            \State Calculating the value of function $f\left( {\left\{ p_{2i - 1}^* \right\}_{i = 1}^{N/2}}  \right)$;
            \State ${q ^{\left( {n + 1} \right)}} = {\rm{EE}}\left( {\left\{ p_{2i - 1}^* \right\}_{i = 1}^{N/2}}  \right)$;
            \State $n = n + 1$;
        \EndWhile
        \Ensure ${\rm{EE}}\left( {\left\{ p_{2i - 1}^* \right\}_{i = 1}^{N/2}} \right)$;
    \end{algorithmic}
\end{algorithm}

\subsection{Finite-alphabet Inputs}
For the finite-alphabet inputs, the achievable rate  expression is given by \eqref{mutual_inf_ACO},
thus, the EE of finite-alphabet inputs ${\rm{E}}{{\rm{E}}_{\rm{F}}}\left( {{\left\{ {{p _{2i - 1}}} \right\}_{i=1}^{N/2}}} \right)$ is given by \eqref{EE_F_long}.
%\begin{align}     \label{EE_F_long}
%& \mathrm{EE}_\mathrm{F}\left( {{\left\{ {{p _{2i - 1}}} \right\}_{i=1}^{N/2}}}\right)  \nonumber\\
%& = \frac{\sum\limits_{i = 1}^{N/2} {{W}\left( { {{\log }_2}M - \frac{1}{{\ln 2}} - \sum\limits_{n = 1}^M {\frac{1}{M}{\mathbb{E}_Z}\left\{ {{{\log }_2}\sum\limits_{k = 1}^M  \exp \left( { - \frac{{{{\left| {\frac{1}{2}{H_{2i - 1}}\sqrt {{p_{2i - 1}}} \left( {{X_{2i - 1,n}} - {X_{2i - 1,k}}} \right) + {Z_{2i - 1}}} \right|}^2}}}{{{\sigma ^2}{W}}}} \right)} \right\}} } \right)}  }{2\sum\limits_{i = 1}^{N/2} p _{2i - 1}  + P_c}.
%\end{align}
\setcounter{equation}{53}

Furthermore,  the average optical power constraint  \eqref{EE_ori_con0} can be restricted to constraint \eqref{MI_con3_jian}.
Thus, the optimal  EE maximization  problem with finite-alphabet inputs under the electrical power constraint, the average optical power constraint, and the
 minimum rate constraint   can be expressed as
 \begin{subequations}    \label{finite_EE_ACO}
 \begin{align}
  \mathop {{\text{maximize}}
 }\limits_{{{\left\{ {{p _{2i - 1}}} \right\}_{i=1}^{N/2}}}} ~& {\rm{E}}{{\rm{E}}_{\rm{F}}}\left( {{\left\{ {{p _{2i - 1}}} \right\}_{i=1}^{N/2}}} \right)  \label{finite_EE_ACO_a}\\
 {\rm{s}}{\rm{.t}}{\rm{.}}{\rm{    }}~& \sum\limits_{i = 1}^{N/2} {{p _{2i - 1}}}  \le \min \left\{ {P,\frac{{4P_o^2}}{{{\mathbb{E}^2}\left\{ {\left| {{X_{2i - 1}}} \right|} \right\}}}} \right\},  \label{finite_EE_ACO_con1}\\
 & \sum\limits_{i = 1}^{N/2} {{R_{{\rm{F}}}}\left( {{ {{p _{2i - 1}}}}} \right)}  \ge {r},  \label{finite_EE_ACO_con2}\\
 & {p _{2i - 1}} \ge 0, \quad i = 1,...,N/2.    \label{finite_EE_ACO_con3}
 \end{align}
 \end{subequations}

Note that there is no closed-form expression for the achievable rate in \eqref{finite_EE_ACO_a},
although ${{R_{{\text{F}}}}\left( \left\{{p _{2i - 1}}\right\} \right)}$ is strictly
concave over its input power.
On the other hand, constraints \eqref{finite_EE_ACO_con1}-\eqref{finite_EE_ACO_con3} form a convex feasible solution set.
Thus, problem \eqref{finite_EE_ACO} is
a concave-linear fractional problem that  can be solved by  Dinkelbach-type algorithms.
      The details are omitted as it   is
similar to the case adopting Gaussian distribution inputs as discussed in Section IV-A.

\subsection{Lower Bound of Mutual Information}
 Note that in the objective function \eqref{finite_EE_ACO_a}, complicated integrals need to be solved with high computational complexity.   {For the same reason as in Section III-C, we exploit the lower bound of achievable rate \eqref{total_bound_ACO} to reduce the  complexity}, and  the corresponding EE  function of  the ACO-OFDM VLC system  is    given by \eqref{EE_ACO}.
% \begin{align}     \label{EE_ACO}
%&{\rm{E}}{{\rm{E}}_{\rm{L}}}\left( {{\left\{ {{p _{2i - 1}}} \right\}_{i=1}^{N/2}}} \right)  \nonumber \\
%&= \frac{{\frac{N{W}}{2}\left( { {{\log }_2}M + 1 - \frac{1}{{\ln 2}}} \right) - \sum\limits_{i = 1}^{N/2} {\sum\limits_{n = 1}^M {\frac{{W}}{M} {{\log }_2}\sum\limits_{k = 1}^M { \exp \left( { - \frac{{{{\left| {\frac{1}{2}{H_{2i - 1}}\sqrt {{p_{2i - 1}}} \left( {{X_{2i - 1,n}} - {X_{2i - 1,k}}} \right)} \right|}^2}}}{{2{\sigma ^2}{W}}}} \right)} } } }}{{2\sum\limits_{i = 1}^{N/2} {{p _{2i - 1}}}  + {P_c}}}.
%\end{align}
\setcounter{equation}{55}

Furthermore,
the   EE problem \eqref{EE_ori}      can be reformulated  as
\begin{subequations}  \label{EE_ACO_lower}
\begin{align}
\mathop {{\text{maximize}}
 }\limits_{{{\left\{ {{p _{2i - 1}}} \right\}_{i=1}^{N/2}}}} ~& {\rm{E}}{{\rm{E}}_{\rm{L}}}\left( {{\left\{ {{p _{2i - 1}}} \right\}_{i=1}^{N/2}}} \right)   \label{EE_lowerbound_ACO}\\
{\rm{s}}{\rm{.t}}{\rm{.}}{\rm{    }}~& \sum\limits_{i = 1}^{N/2} {{p _{2i - 1}}}  \le \min \left\{ {P,\frac{{4P_o^2}}{{{\mathbb{E}^2}\left\{ {\left| {{X_{2i - 1}}} \right|} \right\}}}} \right\},  \label{EE_lowerbound_ACO_con1}\\
 ~& \sum\limits_{i = 1}^{N/2} {{R_{{\rm{L}}}}\left( {{{{p _{2i - 1}}}}} \right)}  \ge {r},  \label{EE_lowerbound_ACO_con2}\\
 ~& {p _{2i - 1}} \ge 0,  \quad i = 1,...,N/2.  \label{EE_lowerbound_ACO_con3}
\end{align}
\end{subequations}

As the lower bound of the  achievable rate \eqref{total_bound_ACO} is concave and differentiable,
problem \eqref{EE_ACO_lower} is  also a concave-linear fractional problem.
At the same time, problem \eqref{EE_ACO_lower} can also be solved by using  Dinkelbach-type algorithms.

\section{Simulation Results and Discussion}

This section presents numerical results to evaluate the proposed  power allocation schemes  for SE  maximization  and EE maximization problems of ACO-OFDM VLC systems.
Consider an indoor ACO-OFDM VLC system installed with  four LEDs, where  a corner of a square room  denotes
the origin  $\left( {0, 0, 0} \right)$ of a three-dimensional Cartesian coordinate
system $\left( {X, Y, Z} \right)$. The location
of receiver is $\left( {0.5, 1, 0} \right)$m, {the locations of  four LEDs are $\left( {1.5, 1.5, 3} \right)$m, $\left( {1.5, 3.5, 3} \right)$m, $\left( {3.5, 1.5, 3} \right)$m, and $\left( {3.5, 3.5, 3} \right)$m, respectively, and the reflection point is $\left(1.5,0,1.5\right)$.
The   basic parameters of the VLC system are listed in Table   \ref{baisc_par}. The channel gain is generated based on the channel model \eqref{channel_model}, \eqref{channel_model_direct}, and \eqref{channel_model_diffuse}\cite{Schulze, Kahn, Jungnickel}.}
Besides, the $R_{\mathrm{L}}\left(p_{2i-1}\right)$ has been added a non-negative real  constant as same as \cite{XiaoLu} in all simulations without the optimality loss of SE- and EE-maximization power allocation.

\begin{table}[htbp]
    \centering
    \caption{Simulation Parameters of  the  ACO-OFDM VLC System.}\label{baisc_par}
    \begin{tabular}{|l|l|}
        \hline
        Definition  & Value   \\
        \hline
        Number of subcarriers, $N$ & $64$   \\
        \hline
        Transmit angle, $\theta$ & ${60^ \circ }$   \\
        \hline
        FOV, $\Psi$ & ${90^ \circ }$   \\
        \hline
        Lambertian emission order, $m$ & 1   \\
        \hline
        Half power angle, ${\Phi _{1/2}}$&${60^ \circ }$ \\
        \hline
        PD collection area, ${A_{\rm{r}}}$ & $1~{\rm{c}}{{\rm{m}}^2}$  \\
        \hline
        Circuit power consumption,  $P_c$   &   $0.2~{\rm{W}}$           \\
        \hline
        Angle of arrival/departure, $\varphi$ & ${45^ \circ }$ \\
        \hline
        Optical filter gain of receiver, $T\left( \varphi  \right)$ & $0~{\rm{dB}}$  \\
        \hline
        Concentrator gain of receiver, $G\left( \varphi  \right)$ & $0~{\rm{dB}}$  \\
        \hline
        Noise PSD, ${\sigma ^2}$ & $10^{-18}~\rm{A^2/Hz}$   \\
        \hline
        Modulation order, ${M}$ & 4-QAM   \\
        \hline
        Bandwidth of each subcarrier, ${W}$ & $1~{\rm{MHz}}$   \\
        \hline
    \end{tabular}
\end{table}

\subsection{Simulation Results of SE Maximization Problem}
In this subsection, we present the results of the proposed three power allocation schemes for  maximizing the SE  for
Gaussian distribution inputs, finite-alphabet inputs, and lower bound  of  the mutual information.

\begin{figure}[htbp]
    \centering
    \includegraphics[height=5.9cm,width=6.32cm]{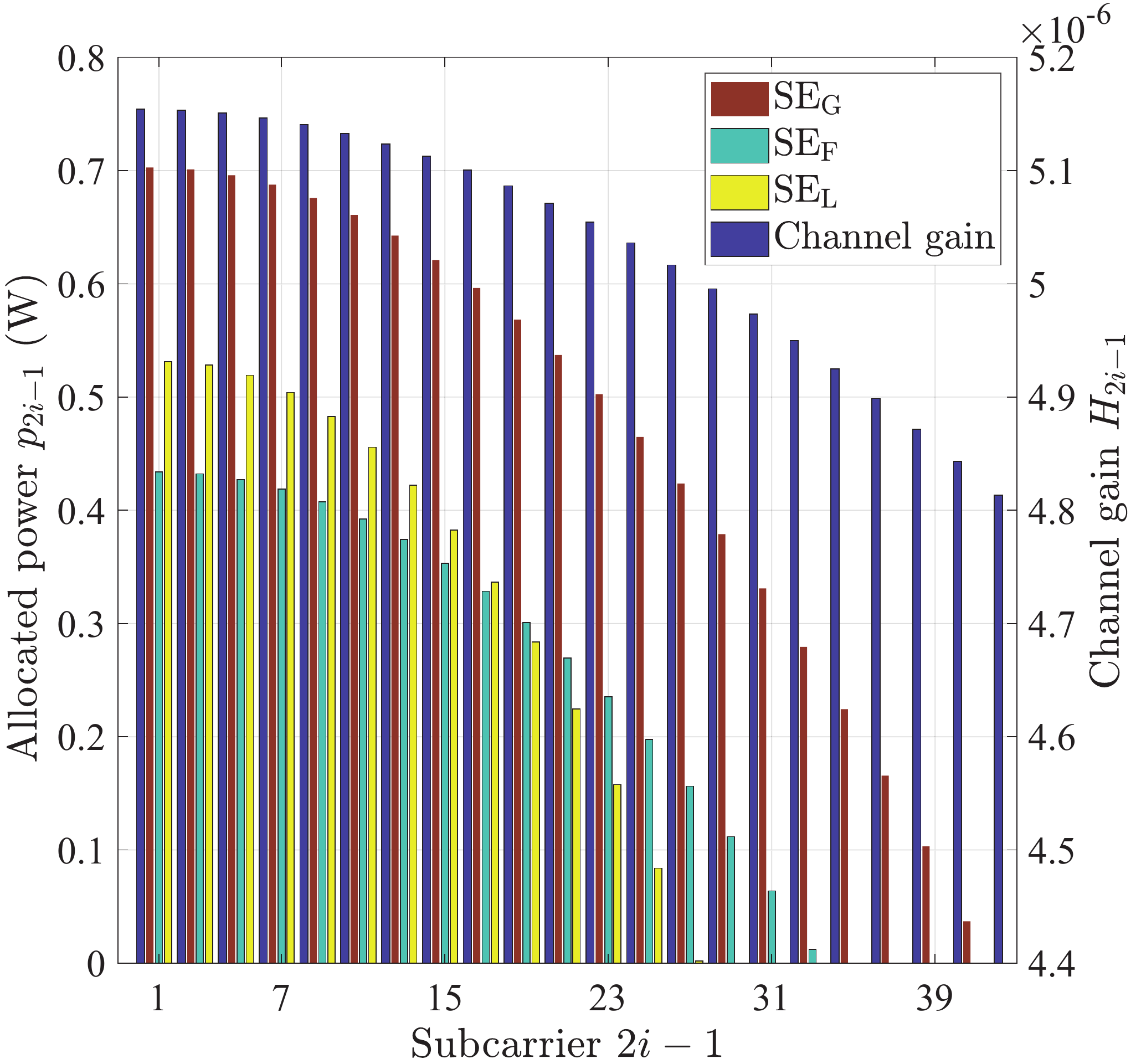}
    \caption{~Allocated power $p_{i}$ versus channel gain $H_{i}$ of  subcarrier $i$ of
        $\text{SE}_\text{G}$, $\text{SE}_\text{F}$, and $\text{SE}_\text{L}$ with ${P} = 20$ W,   ${P_o} = 0.25$ W.}
    \label{fig_channelgain_rate}
\end{figure}

Fig. \ref{fig_channelgain_rate} illustrates allocated power $p_{i}$ versus channel gain $H_{i}$ of  subcarrier $i$ of
$\text{SE}_\text{G}$, $\text{SE}_\text{F}$, and $\text{SE}_\text{L}$, where  ${P} = 20$ (W),   ${P_o} = 0.25$ (W).
As can be observed from   Fig. \ref{fig_channelgain_rate},
the value of the  allocated  power   $p_{i}$ of the $\text{SE}_\text{G}$ is proportional to the corresponding noise level $\frac{{4{\sigma ^2}{W}}}
{{{{\left| {{H_{2i - 1}}} \right|}^2}}}$, which is due to the water-filling solution for the maximization of the SE with Gaussian distribution inputs.
While for the $\text{SE}_\text{F}$ case,
 the allocated power of  $p_{i}$ not only   depends on the noise level, but also depends on the mercury level
$\frac{{4{\sigma ^2}{W}}}{{{{\left| {{H_{2i - 1}}} \right|}^2}}}{G_{2i - 1}}\left( \lambda  \right)$,
 which is due to the mercury-water-filling method for the maximization of the SE with finite-alphabet inputs.
 For  the case of $\text{SE}_\text{L}$,
 the allocated power of each subcarrier is based on $\sum_{n = 1}^M {{{\log }_2}\sum_{k = 1}^M {\exp \left( { - \frac{{{p_{2i - 1}}{{\left| {{H_{2i - 1}}} \right|}^2}{{\left| {{X_{2i - 1,n}} - {X_{2i - 1,k}}} \right|}^2}}}
{{8{\sigma ^2}{W}}}} \right)} } $.
Since $N\pi  \ge \frac{4}{{{\mathbb{E}^2}\left\{ {\left| {{X_{2i - 1}}} \right|} \right\}}}$, we have ${N}\pi P_o^2\ge \frac{{4P_o^2}}{{{\mathbb{E}^2}\left\{ {\left| {{X_{2i - 1}}} \right|} \right\}}}$.
Therefore,  when ${P \ge 2N\pi P_o^2}$,
  the allocated power for Gaussian distribution inputs is  more     than  that for  finite-alphabet inputs.
Moreover, due to the negligible small channel gain, the allocated power from the  $43$-th subcarrier  to the   $63$-th subcarrier   of the  three  cases are zero, which are not shown in
Fig. \ref{fig_channelgain_rate} for brevity.

\begin{figure}[htbp]
    \centering
    \begin{minipage}[b]{0.5\textwidth}
        \centering
        \includegraphics[height=5.9cm,width=6.32cm]{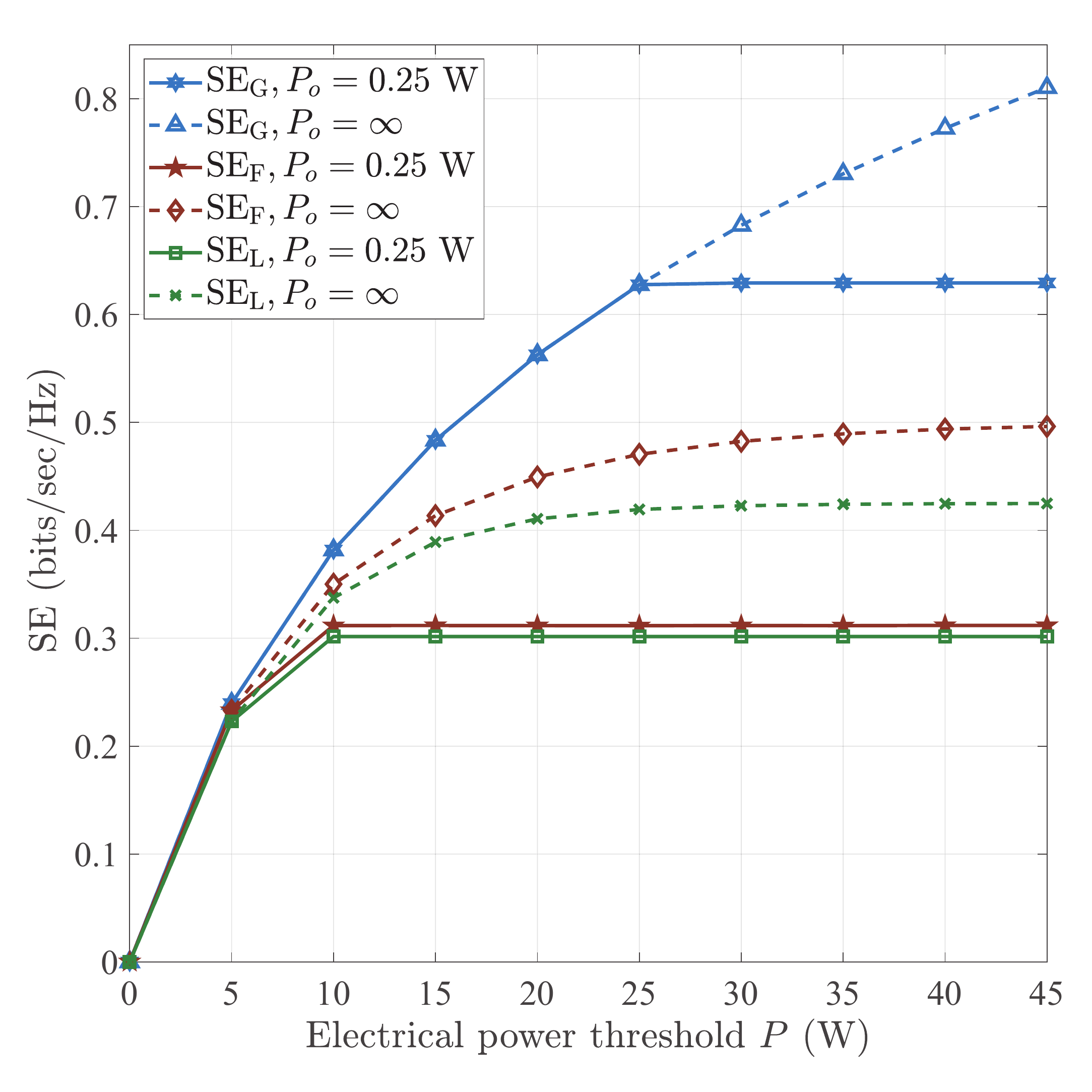}
        \vskip-0.2cm\centering {\footnotesize (a)}
    \end{minipage}
    \begin{minipage}[b]{0.5\textwidth}
        \centering
        \includegraphics[height=5.9cm,width=6.32cm]{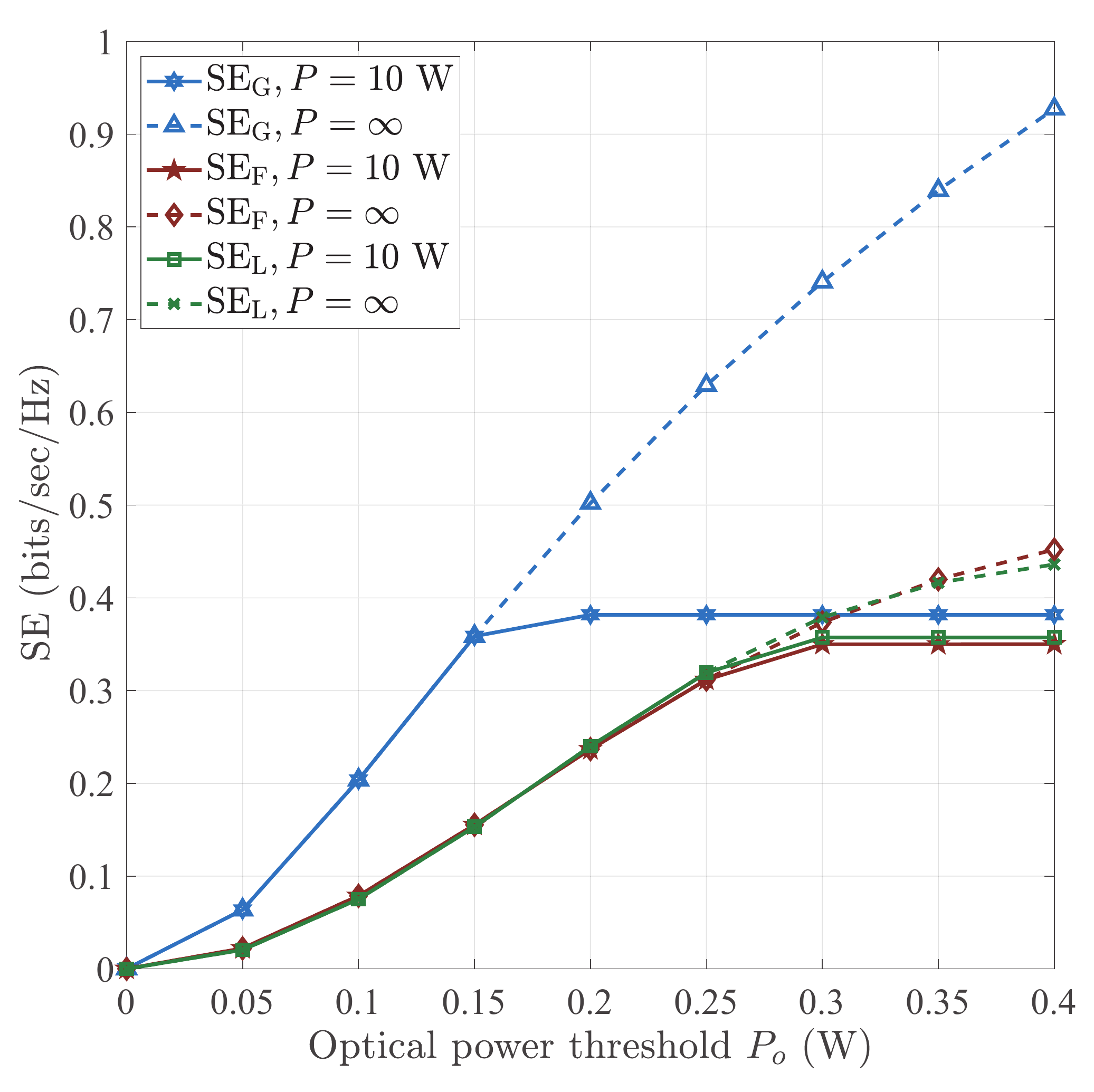}
        \vskip-0.2cm\centering {\footnotesize (b)}
    \end{minipage}
    \caption{ (a) $\text{SE}_\text{G}$, $\text{SE}_\text{F}$, and $\text{SE}_\text{L}$  versus electrical power threshold ${P}$;
        (b)  $\text{SE}_\text{G}$, $\text{SE}_\text{F}$, and $\text{SE}_\text{L}$ versus optical power threshold ${P_{o}}$.}
    \label{fig_SE_P_Po}
\end{figure}

Fig. \ref{fig_SE_P_Po}(a) illustrates $\text{SE}_\text{G}$, $\text{SE}_\text{F}$, and $\text{SE}_\text{L}$ versus the  electrical power threshold
${P}$
with an optical power constraint  ${P_{o}}=0.25 $ (W) and   ${P_{o}}= \infty $ (without optical power constraint), respectively.
As shown in Fig. \ref{fig_SE_P_Po}(a),  for  ${P_{o}}= \infty $ case,
 as ${P}$ increases,   $\text{SE}_\text{G}$ {keeps} increasing, while  $\text{SE}_\text{F}$ and $\text{SE}_\text{L}$ first increase and then remain constant. Moreover, when    ${P}$ is large,  $\text{SE}_\text{G}$ is higher than both $\text{SE}_\text{F}$ and $\text{SE}_\text{L}$.
 This is because the Gaussian distribution can be regarded as a very
high-order constellation modulation input, which is more suitable for high SNRs.
 While for   ${P_{o}}=0.25 $ (W) case, as ${P}$ increases,
 $\text{SE}_\text{G}$, $\text{SE}_\text{F}$, and $\text{SE}_\text{L}$ first increase and then remain constant.
 This is because the allocated power is limited by the optical
power threshold, i.e., ${P_{o}}=0.25 $ (W).
Moreover, it  can be observed from Fig. \ref{fig_SE_P_Po}(a)   that the $\text{SE}_\text{F}$ with finite-alphabet inputs
reaches the maximum point faster than the $\text{SE}_\text{G}$ with Gaussian distribution inputs.
The reason is that the   coefficient of ${P_o}$ of Gaussian distribution inputs  $2{N}\pi $ in   \eqref{simpaco_con1} is larger than that of   finite-alphabet inputs $\frac{4}{{\mathbb{E}^2}\left\{ {\left| {{X_{2i - 1}}} \right|} \right\}}$ in \eqref{power_constr}.
Besides, the gap between  $\text{SE}_\text{F}$ and $\text{SE}_\text{L}$ is small, which means $\text{SE}_\text{L}$ could approximate $\text{SE}_\text{F}$ well enough.

Fig. \ref{fig_SE_P_Po}(b) illustrates  $\text{SE}_\text{G}$, $\text{SE}_\text{F}$, and $\text{SE}_\text{L}$ versus optical power threshold ${P_{o}}$
with an electrical power constraint  ${P}=10 $ (W) and   ${P}= \infty $ (without electrical power constraint), respectively.\footnote{{There might be intersections among $\mathrm{SE}_{\mathrm{L}}$, $\mathrm{SE}_{\mathrm{F}}$, and $\mathrm{SE}_{\mathrm{G}}$ as similar as the Fig. 1, Fig. 2, and Fig. 3 in \cite{XiaoLu}.}}.

It can be seen in Fig. \ref{fig_SE_P_Po}(b), for ${P}= \infty $ case,
 as ${P_{o}}$ increases,
$\text{SE}_\text{G}$ keep increasing,
while $\text{SE}_\text{F}$  and $\text{SE}_\text{L}$ first increase and then remain constant.
Besides,  when   ${P_{o}}$ is large,  $\text{SE}_\text{G}$ is higher than both $\text{SE}_\text{F}$ and $\text{SE}_\text{L}$.
 The reason is  that the Gaussian distribution   is more suitable for high SNRs, which is similar as that in Fig. \ref{fig_SE_P_Po}(a).
 While for ${P}=10 $ (W) case,
$\text{SE}_\text{G}$, $\text{SE}_\text{F}$, and $\text{SE}_\text{L}$
first increase and remain  as a constant.
In fact, the allocated power  is restricted   by the electrical power  constraint, i.e., ${P}=10 $ (W).
When ${P_o}$ is small,
   $\text{SE}_\text{G}$ is higher than $\text{SE}_\text{F}$  and $\text{SE}_\text{L}$. As ${P_o}$  increases, $\text{SE}_\text{G}$  first  reaches its maximum  point, while  $\text{SE}_\text{F}$  and $\text{SE}_\text{L}$  reach to the corresponding saturation points later.
The reason is that
    the coefficient of ${P_o}$ of Gaussian distribution inputs  $2{N}\pi $ in   \eqref{simpaco_con1} is larger than that of   finite-alphabet inputs $\frac{4}{{\mathbb{E}^2}\left\{ {\left| {{X_{2i - 1}}} \right|} \right\}}$ in \eqref{power_constr}.
Thus, with the same  optical power threshold ${P_{o}}$, the Gaussian distribution inputs can allocate more power than that of  finite-alphabet inputs,  which was also verified in Fig. \ref{fig_channelgain_rate}.

\subsection{Simulation Results of EE Maximization Problems}
In this subsection, we present the simulation results for the evaluation of
the EE performance of Gaussian distribution inputs,
finite-alphabet inputs case, and lower bound of mutual information case for ACO-OFDM VLC systems.

\begin{figure}[htbp]
    \centering
    \includegraphics[height=5.9cm,width=6.32cm]{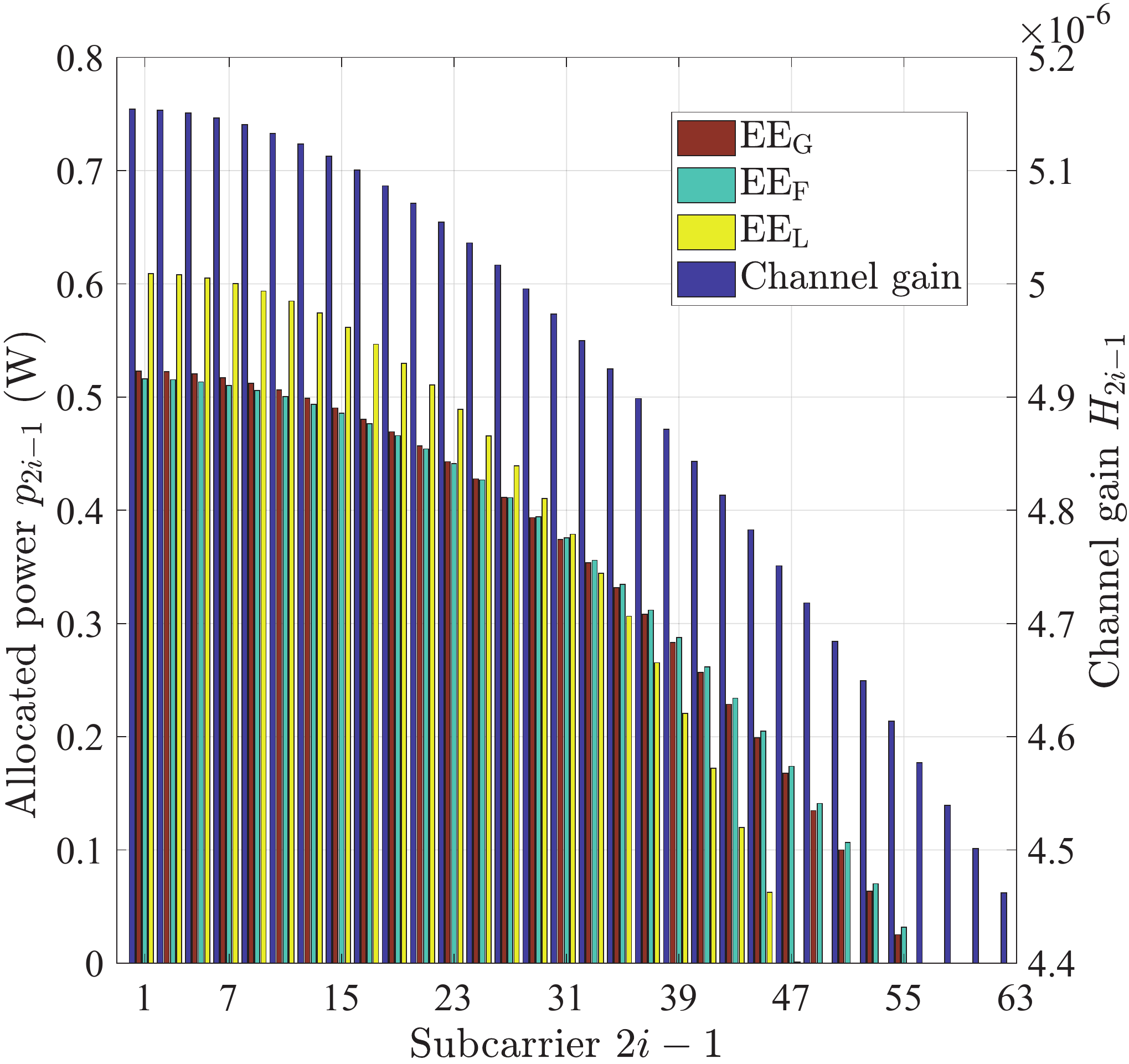}
    \caption{~Allocated power $p_{i}$ versus channel gain $H_{i}$ of  subcarrier $i$ of $\text{EE}_\text{G}$, $\text{EE}_\text{F}$, and $\text{EE}_\text{L}$ with ${P} = 20$ W,   ${P_o} = 0.25$ W.}
    \label{fig_EE_channel}
\end{figure}

Fig. \ref{fig_EE_channel} illustrates the   different
allocated power $p_{i}$ of ${\text{EE}}_\text{G}$, ${\text{EE}}_\text{F}$, and ${\text{EE}_\text{L}}$ versus channel gain $H_{i}$ of  subcarrier $i$ respectively, where  ${P} = 20$ (W),   ${P_o} = 0.25$ (W), and  $r = 1$ (bits/sec/Hz).
  From Fig. \ref{fig_EE_channel}, we can see that    the allocated power  of subcarrier $i$  of   ${\text{EE}}_\text{G}$   is   proportional to its channel gain, which is due to the power allocation strategy in  \eqref{ee_Gau_alpha} and \eqref{ee_Gau_alpha_P}. While  the allocated power  of subcarrier $i$  of both ${\text{EE}}_\text{F}$ and ${\text{EE}}_\text{L}$  depend on
      both channel gains and   MMSE functions.
  Compared with   SE maximization problems,
     both the objective function and   rate constraints of the EE maximization problems    are different,
     which accounts for the different  power allocation in Fig. \ref{fig_EE_channel}
     compared with that in Fig. \ref{fig_channelgain_rate}.

\begin{figure}[htbp]
    \centering
    \begin{minipage}[b]{0.5\textwidth}
        \centering
        \includegraphics[height=5.9cm,width=6.32cm]{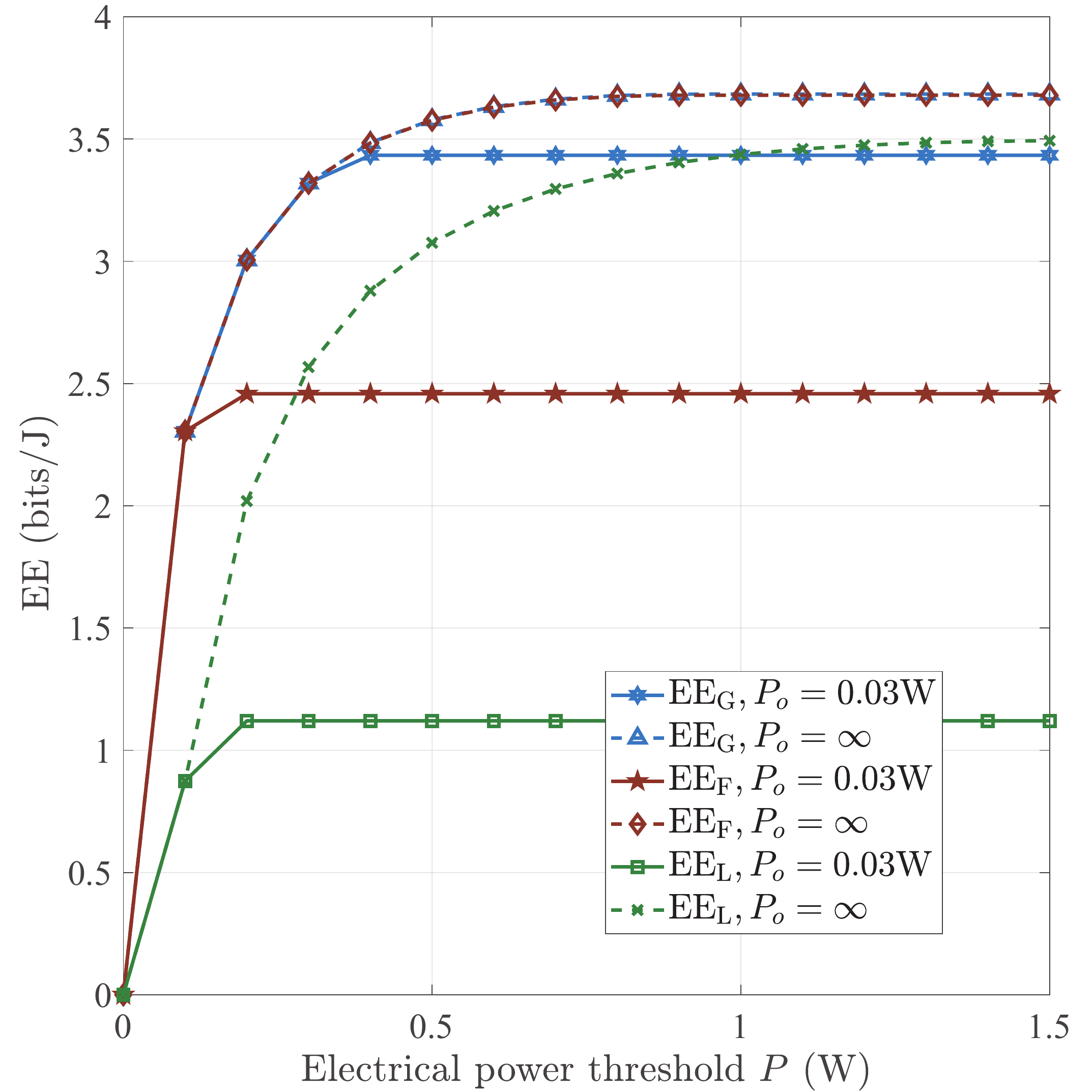}
        \vskip-0.2cm\centering {\footnotesize (a)}
    \end{minipage}
    \begin{minipage}[b]{0.5\textwidth}
        \centering
        \includegraphics[height=5.9cm,width=6.32cm]{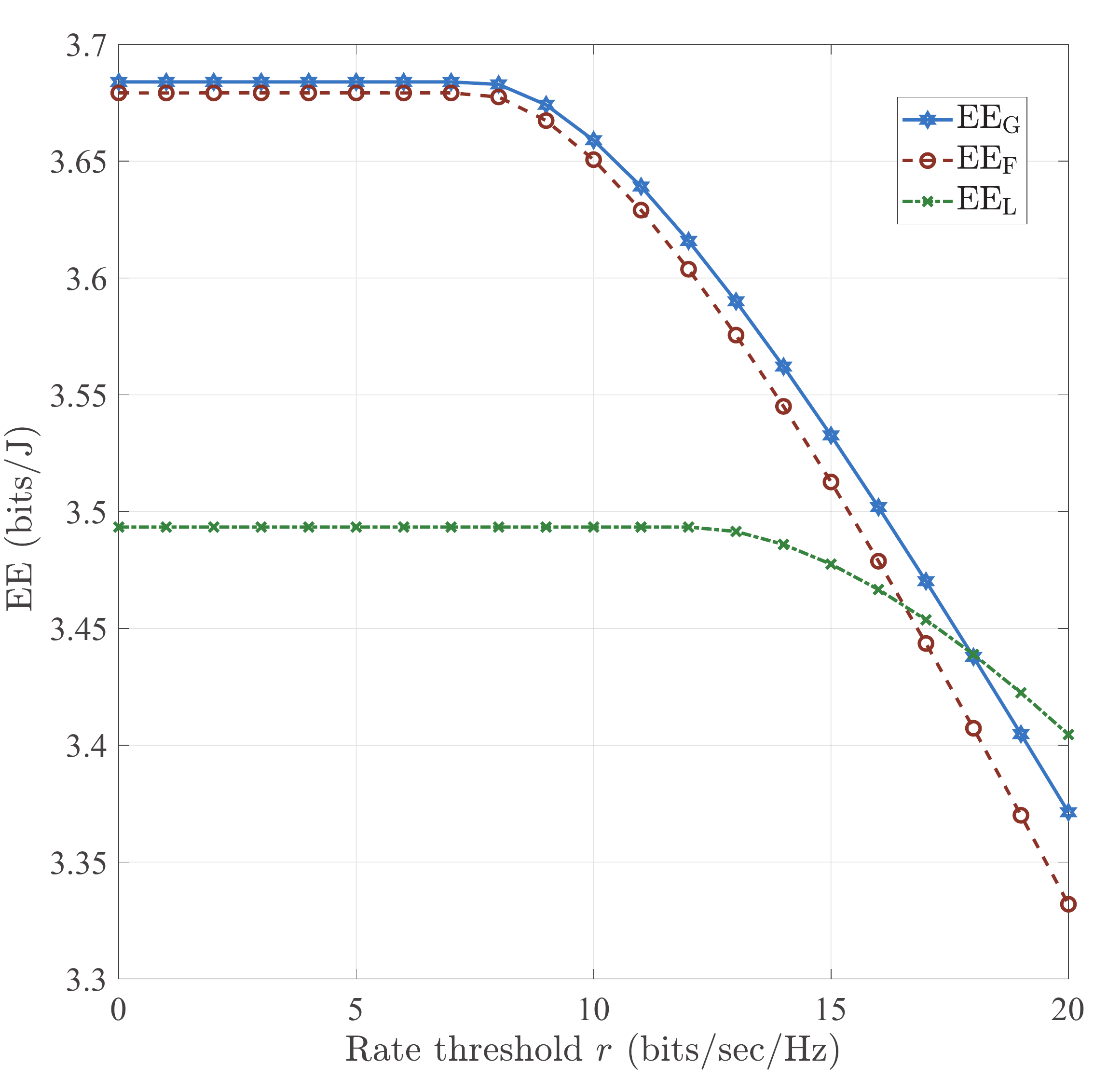}
        \vskip-0.2cm\centering {\footnotesize (b)}
    \end{minipage}
    \caption{ (a) ${\text{EE}}_\text{G}$, ${\text{EE}}_\text{F}$, and ${\text{EE}_\text{L}}$ versus electrical power threshold ${P}$ with rate constraint $r= 0.1$ bits/sec/Hz and two different optical power thresholds  ${P_o} = 0.03$ W,  and  ${P_o} = \infty$;
        (b)  ${\text{EE}}_\text{G}$, ${\text{EE}}_\text{F}$, and ${\text{EE}_\text{L}}$ versus  rate threshold $r$  with   power threshold ${P} = 20$ W and optical power threshold ${P_o} = 1$ W.}
    \label{fig_EE_P_Rate}
\end{figure}

Fig. \ref{fig_EE_P_Rate}(a) depicts ${\text{EE}}_\text{G}$, ${\text{EE}}_\text{F}$, and ${\text{EE}_\text{L}}$ versus electrical power threshold ${P}$ with an  optical power threshold ${P_o} = 0.03$ (W) and ${P_o} = \infty$ (without optical power constraint) respectively, where the rate constraint $r= 0.1$ (bits/sec/Hz).We see from Fig. \ref{fig_EE_P_Rate}(a)   that for ${P_o} = \infty$ case, as ${P}$ increases,
 ${\text{EE}}_\text{G}$, ${\text{EE}}_\text{F}$, and ${\text{EE}_\text{L}}$  first increase and then remain constant.
 This is because the  optimal EE  remains a constant  when it has reached the maximum value.
 Moreover,  the value of  ${\text{EE}}_\text{F}$ approaches  to  that of ${\text{EE}}_\text{G}$, which are higher than the value of ${\text{EE}_\text{L}}$.
 While for ${P_o} = 0.03$ case,
as ${P}$ increases, the ${\text{EE}}_\text{G}$, ${\text{EE}}_\text{F}$, and ${\text{EE}_\text{L}}$  first increase and then remain constant.
The reason is that ${\text{EE}}_\text{G}$, ${\text{EE}}_\text{F}$, and ${\text{EE}_\text{L}}$ is limited by the  optical power constraint ${P_o} = 0.03$ (W).  Besides, for  the large ${P}$,  the value of  ${\text{EE}}_\text{G}$ are the highest of the three power allocation schemes, while  ${\text{EE}}_\text{L}$ is the lowest.

Fig. \ref{fig_EE_P_Rate}(b) depicts ${\text{EE}}_\text{G}$, ${\text{EE}}_\text{F}$, and ${\text{EE}_\text{L}}$ versus  rate threshold $r$  with   power threshold ${P} = 20$ (W) and optical power threshold ${P_o} = 1$ (W).
As shown in Fig. \ref{fig_EE_P_Rate}(b),   ${\text{EE}}_\text{G}$ is higher than  ${\text{EE}}_\text{F}$ and ${\text{EE}}_\text{L}$,
and the gap between  ${\text{EE}}_\text{G}$ and  ${\text{EE}}_\text{F}$ is small.
 Moreover, the EE  of three cases first remains constant and then decreases,
as the rate threshold $r$ increases. Indeed,
when the value of the rate threshold $r$ is small, the performed power allocation can easier in satisfying the rate requirement and thus the ${\text{EE}}$ does not change.
While for a high rate threshold  $r$,
 the resource allocation in the system becomes less feasible in allocating power
as it is forced to consume more power to
satisfy the stringant rate constraint,
 and therefore  the optimal EE decreases.
Moreover, the gap between  ${\text{EE}}_\text{F}$  and ${\text{EE}_\text{L}}$
      increases as rate threshold ${r}$ increases\footnote{{There might be intersections among $\mathrm{EE}_{\mathrm{L}}$, $\mathrm{EE}_{\mathrm{F}}$, and $\mathrm{EE}_{\mathrm{G}}$ as similar as  Fig. 1, Fig. 2, and Fig. 3 in \cite{XiaoLu}.}}.

\subsection{Relationship Between SE and EE}

\begin{figure}[htbp]
    \centering
    \begin{minipage}[b]{0.5\textwidth}
        \centering
        \includegraphics[height=5.9cm,width=6.32cm]{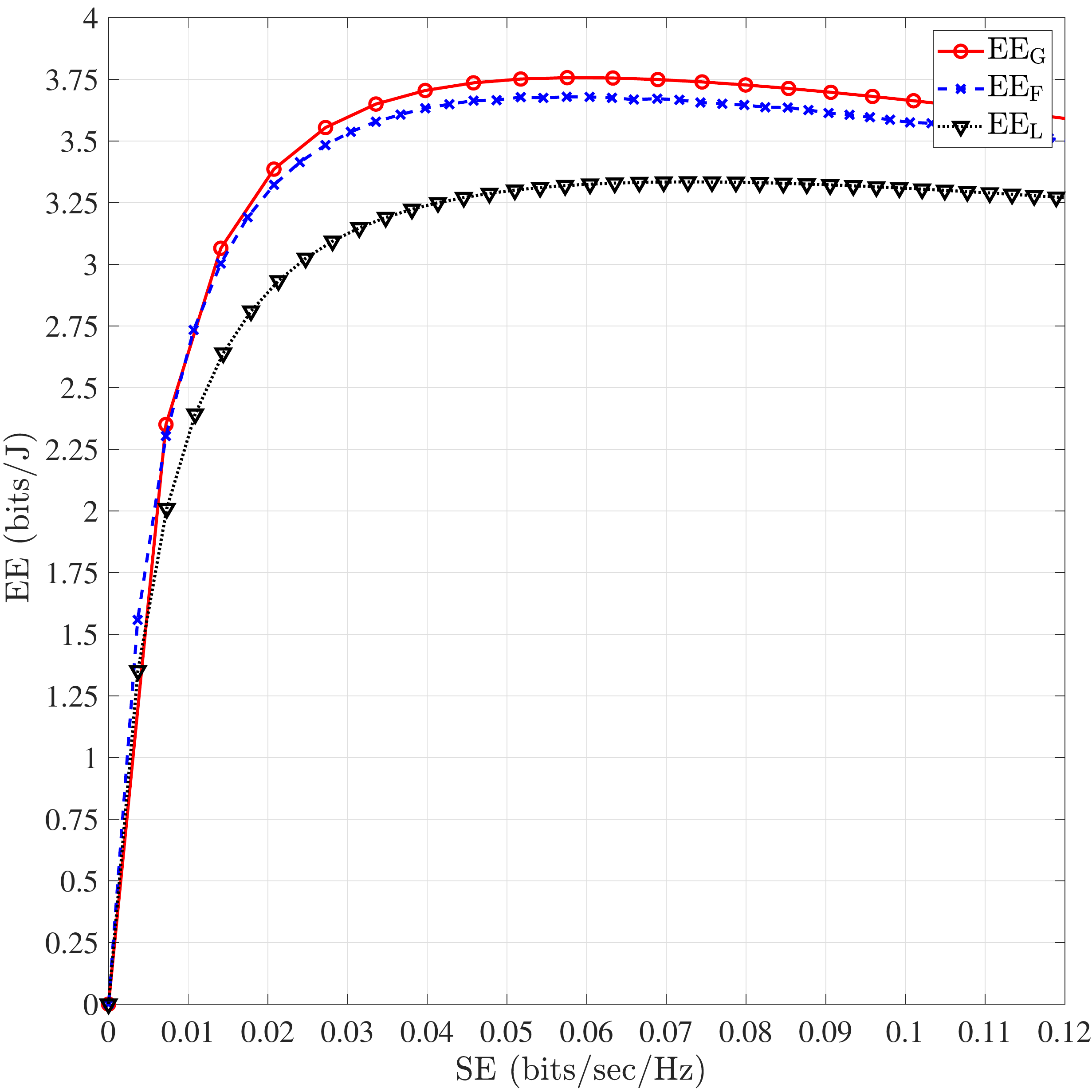}
        \vskip-0.2cm\centering {\footnotesize (a)}
    \end{minipage}
    \begin{minipage}[b]{0.5\textwidth}
        \centering
        \includegraphics[height=5.9cm,width=6.32cm]{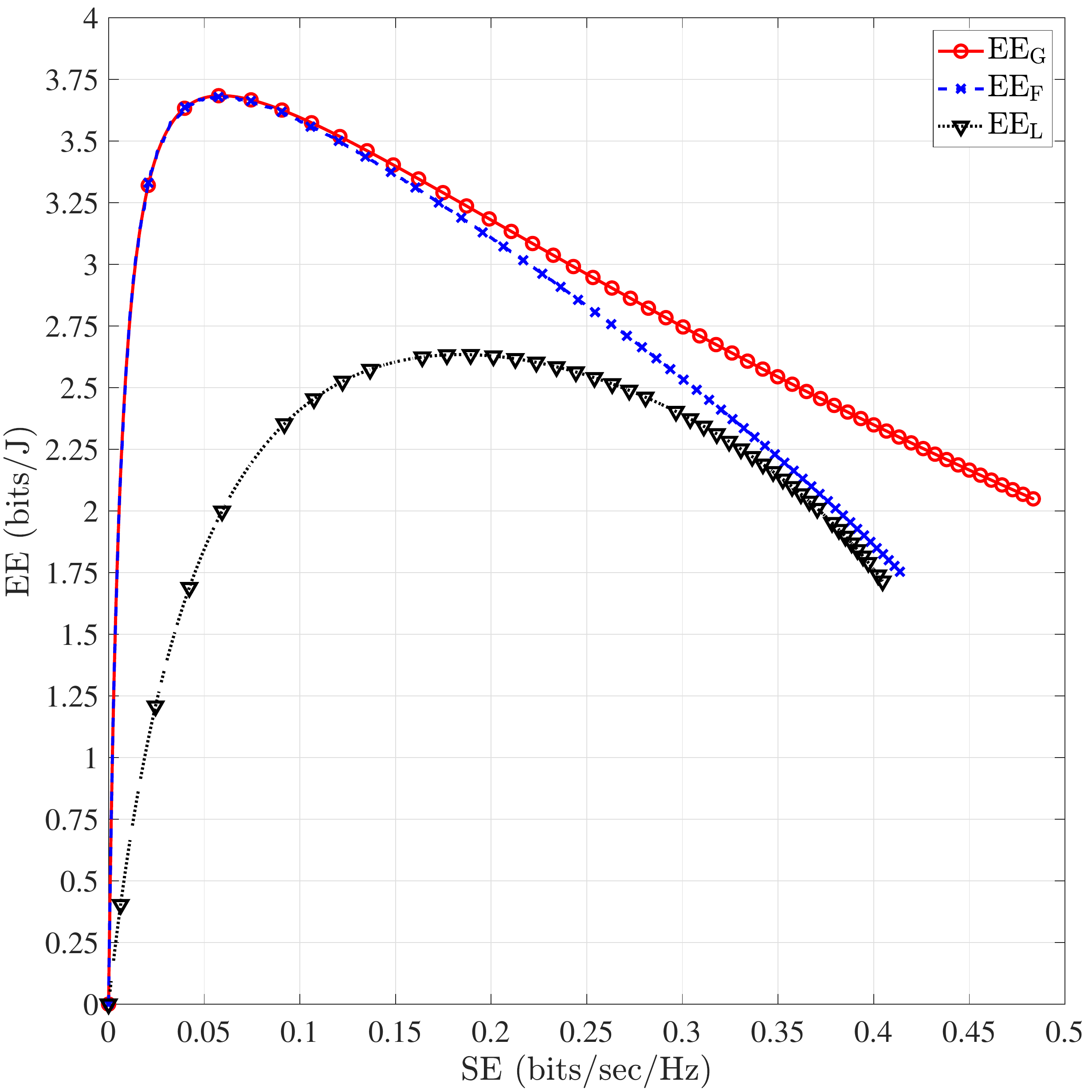}
        \vskip-0.2cm\centering {\footnotesize (b)}
    \end{minipage}
    \caption{ (a) ${\text{EE}}_\text{G}$, ${\text{EE}}_\text{F}$, and ${\text{EE}_\text{L}}$ versus the ${\text{SE}}_\text{G}$, ${\text{SE}}_\text{F}$, and ${\text{SE}_\text{L}}$ with optical power threshold  ${P_{o}}=0.03 $ W;
        (b)  ${\text{EE}}_\text{G}$, ${\text{EE}}_\text{F}$, and ${\text{EE}_\text{L}}$ versus the ${\text{SE}}_\text{G}$, ${\text{SE}}_\text{F}$, and ${\text{SE}_\text{L}}$ with optical power threshold  ${P_{o}}=\infty $.}
    \label{fig_EE_SE}
\end{figure}

To guarantee the QoS to users with affordable energy,
EE and SE in a specific system are used to evaluate the
performances of energy and spectral usage.
Especially, for achieving a good  balance performance of VLC equipments,
the
tradeoff between SE and EE should be delicately considered.
Based on \eqref{se_Gau_1121} and \eqref{ee_Gau_1121},
the relationship between SE and EE
of Gaussian distribution
for ACO-OFDM can be discussed as
\begin{align}      \label{EE_SE_Gau}
{\rm{E}}{{\rm{E}}_{\rm{G}}}\left( {{\left\{ {{p _{2i - 1}}} \right\}_{i=1}^{N/2}}} \right) = \frac{2NW}{{2\sum\limits_{i = 1}^{N/2} {{p _{2i - 1}}}  + {P_c}}}{\rm{S}}{{\rm{E}}_{\rm{G}}}\left( {{\left\{ {{p _{2i - 1}}} \right\}_{i=1}^{N/2}}} \right).
\end{align}
Similarly, the relationship between EE and SE
of finite-alphabet inputs and the lower bound of mutual information
for ACO-OFDM are similar to that of the Gaussian distribution.

Fig. \ref{fig_EE_SE}(a) shows the ${\text{EE}}_\text{G}$, ${\text{EE}}_\text{F}$, and ${\text{EE}_\text{L}}$ versus the ${\text{SE}}_\text{G}$, ${\text{SE}}_\text{F}$ and ${\text{SE}_\text{L}}$ with optical power threshold  ${P_{o}}=0.03 $ (W).
It can be seen from Fig. \ref{fig_EE_SE} that there is a non-trivial tradeoff between the system SE and EE. In practice, as ${\text{SE}}$ increases, the ${\text{EE}}$ increases at first and then decreases. In particular, there exists an optimal SE to maximize EE.
 Moreover, the peak of ${\text{EE}}_\text{G}$ is the highest while the maximum ${\text{EE}}_\text{F}$ is higher than that of ${\text{EE}_\text{L}}$, and this phenomenon was also verified in
 Fig. \ref{fig_EE_P_Rate} (a) and (b).

{Fig. \ref{fig_EE_SE}(b) depicts the  tradeoff  between the  SE and the EE with the optical power threshold $P_o=\infty$. It can be observed that, at the low SE region,
   $\mathrm{EE}_{\mathrm{G}}$ is close to $\mathrm{EE}_{\mathrm{F}}$, and       $\mathrm{EE}_{\mathrm{F}}$ is close to $\mathrm{EE}_{\mathrm{L}}$ at the high SE region.}

\section{Conclusion}
In this study, we addressed the problem of designing optimal power allocation schemes to maximize the SE and the  EE of ACO-OFDM in VLC systems with Gaussian  distributions  and   finite-alphabet inputs.
We first derived the achievable rates and the average optical power constraint for ACO-OFDM VLC systems  with  the above two mentioned inputs. Then,   we derived the    optimal power allocation schemes to maximize the
SE of ACO-OFDM systems.
Specifically, for Gaussian distribution inputs,   the water-filling-based  power allocation scheme  was presented  to  maximize the SE.
By exploiting    the relationship between the mutual information and MMSE, the optimal power
allocation scheme was derived   to  maximum the SE
 system with finite-alphabet inputs.
 We further developed the optimal power allocation scheme to maximize
  the EE      of ACO-OFDM VLC  systems  with     Gaussian  distributions  and finite-alphabet inputs.  By adopting Dinkelbach-type algorithm, the EE   maximization problems were   transformed into   convex problems
 and the interior point algorithm was exploited to obtain the optimal solution.
 Besides,  to reduce the computational
complexity of finite-alphabet inputs cases, we derived    the  closed-form lower bounds of mutual information for both the SE and the EE maximization problems.
 Finally, we showed the relationship between
 the SE and the EE      of ACO-OFDM VLC  systems.

\bibliographystyle{IEEE-unsorted}
\bibliography{refs0308}

\end{document}